\documentclass[11pt]{article}
\usepackage{authblk}
\usepackage[left=1in,right=1in,top=1in,bottom=1in]{geometry}
\usepackage{empheq}
\usepackage{amssymb}
\usepackage{dsfont}
\usepackage[pdfstartview=FitH,pdfpagemode=None]{hyperref}
\usepackage{mathtools}
\usepackage{youngtab}
\usepackage{cite}
\begin{document}
\title{Antisymmetric Wilson loops in $\mathcal{N}=4$ SYM: from exact results to non-planar corrections}
\author[1]{Anthonny F. Canazas Garay}
\affil[1]{Instituto de F\'isica, Pontificia Universidad Cat\'olica de Chile \authorcr
Casilla 306, Santiago, Chile}
\author[2]{Alberto Faraggi}
\affil[2]{Departamento de Ciencias F\'isicas, Facultad de Ciencias Exactas, Universidad Andr\'es Bello \authorcr Sazie 2212, Piso 7, Santiago, Chile}
\author[3,4]{Wolfgang M\"uck}
\affil[3]{Dipartimento di Fisica ``Ettore Pancini", Universit\`a degli Studi di Napoli ``Federico II" \authorcr Via Cintia, 80126 Napoli, Italy}
\affil[4]{Istituto Nazionale di Fisica Nucleare, Sezione di Napoli \authorcr Via Cintia, 80126 Napoli, Italy}
\date{}
\maketitle
\begin{abstract}
We consider the vacuum expectation values of $1/2$-BPS circular Wilson loops in $\mathcal{N}=4$ super Yang-Mills theory in the totally antisymmetric representation of the gauge group $U(N)$ or $SU(N)$. Localization and matrix model techniques provide exact, but rather formal, expressions for these expectation values. In this paper we show how to extract the leading and sub-leading behavior in a $1/N$ expansion with fixed 't~Hooft coupling starting from these exact results. This is done by exploiting the relation between the generating function of antisymmetric Wilson loops and a finite-dimensional quantum system known as the truncated harmonic oscillator. Sum and integral representations for the $1/N$ terms are provided.
\end{abstract}
\newpage
\tableofcontents
%

\numberwithin{equation}{section}

\newcommand{\ie}{i.e.,\ }
\newcommand{\eg}{e.g.,\ }

\newcommand{\const}{\operatorname{const.}}


\newcommand{\rmd}{\,\mathrm{d}}

\newcommand{\Tr}{\operatorname{Tr}}

\newcommand{\idmat}{\mathbb{I}}
\newcommand{\Itwo}{\idmat_2}
\newcommand{\Ifour}{\idmat_4}

\newcommand{\flatind}[1]{{\underline{#1}}}

\newcommand{\re}{\operatorname{Re}}
\newcommand{\im}{\operatorname{Im}}

\newcommand{\e}[1]{\operatorname{e}^{#1}}

\newcommand{\hypF}[1]{\operatorname{F}\left(#1\right)}
\newcommand{\hypFApp}[2]{\operatorname{F_{#1}}\left(#2\right)}
\newcommand{\B}[1]{\operatorname{B}\left(#1\right)}
\newcommand{\Hl}[1]{\operatorname{Hl}\left(#1\right)}
\newcommand{\Laguerre}{\operatorname{L}}
\newcommand{\ChebU}{\operatorname{U}}
\newcommand{\WhitM}{\operatorname{M}}
\newcommand{\Hermite}{\operatorname{H}}
\newcommand{\He}{\operatorname{He}}
\newcommand{\BesselI}[1][0]{\operatorname{I}_{#1}}

\newcommand{\op}{\mathcal{O}}

\newcommand{\vev}[1]{\left\langle #1 \right\rangle}

\newcommand{\ZZ}{\mathcal{Z}}
\newcommand{\tZZ}{\tilde{\mathcal{Z}}(u)}
\newcommand{\FF}[1][]{\mathcal{F}_{#1}}
\newcommand{\tFF}[1][]{\tilde{\mathcal{F}}_{#1}(u)}

\newcommand{\N}{\mathcal{N}}

\newcommand{\Order}{\mathcal{O}}

\newcommand{\unit}[1]{\operatorname{#1}}

\newcommand{\bra}[1]{\langle #1 |}
\newcommand{\ket}[1]{| #1 \rangle}
\newcommand{\braket}[2]{\langle #1 | #2 \rangle}

\newcommand{\ths}{\theta_\ast}
\section{Introduction}\label{Sec:Intro}
Wilson loop operators have played a central role in the development of gauge/gravity dualities \cite{Maldacena:1998im,Rey:1998ik}. In this context, the $\frac{1}{2}$-BPS circular loop in $\mathcal{N}=4$ super Yang-Mills theory with $U(N)$ and $SU(N)$ gauge groups and 't Hooft coupling $\lambda$ has received special attention, mainly due to the conjecture put forward in \cite{Erickson:2000af,Drukker:2000rr} that its expectation value is captured exactly, to all orders in $N$ and $\lambda$, by a Gaussian matrix model. This conjecture was later proved in \cite{Pestun:2007rz} using supersymmetric localization, a technique that has since provided theorists with a several other exact results in supersymmetric gauge theories. 
These include, for example, Wilson loops preserving less supersymmetry \cite{Zarembo:2002an,Drukker:2006ga,Drukker:2007yx,Drukker:2007dw,Drukker:2007qr}, correlators of Wilson loops with chiral primary operators \cite{Semenoff:2006am,Giombi:2006de,Gomis:2008qa,Giombi:2009ds,Bassetto:2009rt,Bassetto:2009ms,Giombi:2012ep,Bonini:2014vta}, correlators between Wilson loops  \cite{Aguilera-Damia:2017znn}, as well as Wilson loops and their correlators in $\mathcal{N}=2$ super Yang-Mills theory \cite{Fraser:2011qa,Russo:2012ay,Russo:2013sba,Fraser:2015xha,Liu:2017fiq,Russo:2017ngf,Billo:2018oog}.


A key ingredient in the description of Wilson loop operators is the representation of the gauge group, typical ones for $U(N)$ and $SU(N)$ being the fundamental, totally symmetric and totally antisymmetric representations. Depending on the rank of the representation, the holographic dual corresponds to probe strings or D-branes \cite{Gomis:2006sb,Yamaguchi:2006tq} propagating on $AdS_5\times S^5$, or to fully back-reacted bubbling geometries \cite{Yamaguchi:2006te,Lunin:2006xr,DHoker:2007mci,Okuda:2008px}. In all cases, the on-shell action of the gravitational object agrees perfectly with the matrix model calculation \cite{Hartnoll:2006is} at leading order in $1/N$ and $1/\sqrt{\lambda}$. We expect, however, that a thorough examination should yield a match at the next-to-leading order as well. It is then crucial to systematically extract these corrections on both sides of the duality.

From the gravitational perspective the calculation of sub-leading corrections amounts to analyzing semi-classical fluctuations of the background configurations and computing the corresponding one-loop partition functions. There has been considerable effort  in this direction \cite{Forste:1999qn,Drukker:2000ep,Kruczenski:2008zk,Faraggi:2011bb,Faraggi:2011ge,Faraggi:2014tna,Forini:2015bgo,Faraggi:2016ekd,Forini:2017whz,Aguilera-Damia:2018bam,Aguilera-Damia:2018twq}, although the calculations are plagued with ambiguities inherent to string theory in curved spaces. 

On the gauge theory side, solving the matrix model is a non-trivial task, and computing sub-leading corrections is a conceptually clear, albeit technically difficult procedure. Recently, sub-leading corrections in $1/N$ to the Wilson loop in the totally antisymmetric representation have been found using loop equation techniques \cite{Gordon:2017dvy} and topological recursion in the Gausssian matrix model \cite{Okuyama:2017feo} (see also \cite{Chen-Lin:2016kkk} for the case of symmetric representations). A particularly interesting development was reported in \cite{Fiol:2013hna}, where the authors managed to compute the exact vacuum expectation value of the circular Wilson loop in arbitrary irreducible representations using the method of orthogonal polynomials. In other words, they solved the matrix model exactly. While useful for some purposes, their results are rather formal, and it is not at all obvious how to extract the large-$N$ (and large-$\lambda$) limit from them, much less any sub-leading corrections, except for simple cases. So far, they have only been used for numerical comparison with other approaches.  

In this paper, we address and solve the problem of extracting the leading and first sub-leading terms in the $1/N$ expansion of the generating function of totally antisymmetric Wilson loops starting from the exact results in \cite{Fiol:2013hna}. Our motivation is two-fold. First, it is obviously interesting to see how it can be done. Second, the techniques used in \cite{Gordon:2017dvy,Okuyama:2017feo} solve the matrix model in the continuum limit, in which the discrete poles of the matrix model resolvent give rise to a cut singularity and a continuous eigenvalue density. In other words, the analyticity properties of the resolvent change in this limit, and it would be interesting to see whether and where this has any implications. We anticipate that we do not find any at order $1/N$. 

The paper is organized as follows. In section~\ref{Sec:GMM}, we review the Gaussian matrix model and reproduce the formal result of \cite{Fiol:2013hna} for the generating function of antisymmetric Wilson loops. At the same time, we establish a connection with the algebra of the truncated harmonic oscillator, which will be crucial for our developments. In section~\ref{WL0}, we extract the leading large-$N$ behaviour (fixed $\lambda$) of the generating function from the formal solution of the matrix model. The most substantial part of the paper is section~\ref{WL1}, which contains the calculation of the $1/N$ terms of the Wilson loop generating function, both for general $\lambda$ and in the holographic regime of large $\lambda$. Our results are shown to agree with \cite{Gordon:2017dvy,Okuyama:2017feo}, but turn out to be somewhat simpler in their final form. Section~\ref{concs} contains the conclusions. Some technical details are deferred to the appendices.

\section{Gaussian Matrix Model}
\label{Sec:GMM}
Localization techniques \cite{Pestun:2007rz} map the expectation value of the circular Wilson loop in $\mathcal{N}=4$ SYM with gauge group $U(N)$ to an expectation value in a Gaussian matrix model. We begin this section by reviewing this model and developing some results that will be relevant for what follows.
\subsection{Partition Function and Expectation Values}
The Gaussian matrix model is defined by the partition function
\begin{empheq}{equation}
	Z =\int[dX]\exp\left(-\frac{2N}{\lambda}\textrm{Tr} \left(X^2\right) \right)\,,
\end{empheq}
where $X$ is a $N\times N$ hermitian matrix. If the gauge group is $SU(N)$ the matrices are also traceless, condition that can be implemented with a Lagrange multiplier. The expectation value of any quantity $F(X)$ in the Hermitian ensemble is then given by
\begin{empheq}{equation}\label{MM:vevs}
	\vev{F(X)} =\frac{1}{Z}\int[dX]\,F(X)\exp\left(-\frac{2N}{\lambda}\textrm{Tr}\left(X^2\right)\right)\,.
\end{empheq}
When $F$ is invariant under similarity transformations, i.e., $F(VXV^{-1})=F(X)$, one can diagonalize the matrix $X$ in terms of its eigenvalues $x_n$, $n=1,\ldots,N$, and integrate out the remaining ``angular'' variables. This yields
\begin{empheq}{alignat=7}
	\vev{F(X)} =\frac{1}{Z} \int\left[\prod_{m=1}^N dx_m\right]\Delta^2(x)F(x)\exp\left(-\frac{2N}{\lambda}\sum_{n=1}^Nx_n^2\right)\,,
\end{empheq}
where the transformation Jacobian
\begin{empheq}{equation}
	\Delta(x)=\det \left[x_m^{n-1}\right]=
	\left|
	\begin{array}{ccccc}
		1 & x_1 & x_1^2 & \cdots & x_1^{N-1}
		\\
		1 & x_2 & x_2^2 & \cdots & x_2^{N-1}
		\\
		\vdots & \vdots & \vdots & \ddots & \vdots
		\\
		1 & x_N & x_N^2 & \cdots & x_N^{N-1}
	\end{array}
	\right|
\end{empheq}
is known as the Vandermonde determinant. 

The Vandermonde determinant enjoys several properties that we can exploit to our advantage. For example, one can show that
\begin{empheq}{equation}\label{MM:Delta.prop}
	\Delta(x) =\prod_{1\leq m<n\leq N}|x_n-x_m|
	\qquad\Rightarrow\qquad
	\Delta(x+y)=\Delta(x)\,,
	\qquad
	\Delta(\alpha x)=\alpha^{\frac{N(N-1)}{2}}\Delta(x)\,,
\end{empheq}
so $\Delta(x)$ is invariant under uniform translations and transforms simply under rescalings. Another key observation is that, up to an overall constant, the Vandermonde determinant is the same as
\begin{empheq}{equation}\label{MM:Delta}
	\tilde{\Delta}(x)=\textrm{det}\left[P_{n-1}(x_m)\right]=C_N\Delta(x)\,,
	\qquad
	C_N=\prod_{n=0}^{N-1}h_n\,,
\end{empheq}
where $P_n(x)= h_n x^n +\cdots$ is any family of polynomials that can be chosen to our convenience. Using these facts, the partition function may be written as
\begin{empheq}{equation}
	Z = C_N^{-2}g^{N^2}\int\left[\prod_{m=1}^Ndx_m\right]\tilde{\Delta}(x+y)\tilde{\Delta}(x+z)\exp\left(-\frac{1}{2}\sum_{n=1}^Nx_n^2\right)\,.
\end{empheq}
Here, $y$ and $z$ are arbitrary constants, and the coupling $g$ is defined by
\begin{empheq}{equation}
	g=\sqrt{\frac{\lambda}{4N}}\,.
\end{empheq}

We now introduce a quantity that will play a central role in our analysis of the circular Wilson loop. Let us define the $N\times N$ matrix
\begin{empheq}{equation}\label{MM:I.def}
	I_{mn}(y,z)=\int_{-\infty}^{\infty}dx\,P_{m-1}(x+y)P_{n-1}(x+z)e^{-\frac{1}{2}x^2}\,.
\end{empheq}
Although this definition might seem spurious at this point, its relevance will become clear momentarily. Writing the determinant \eqref{MM:Delta} explicitly as
\begin{empheq}{equation}\label{MM:Deltat.def}
	\tilde{\Delta}(x)=\sum_{m_1,\,\ldots,\,m_N}\epsilon^{m_1\cdots m_N}P_{m_1-1}(x_1)\cdots P_{m_N-1}(x_N)\,,
\end{empheq}
we find that
\begin{empheq}{equation}
	Z=C_N^{-2}g^{N^2}N!\,\textrm{det}\left[I(y,z)\right]\,.
\end{empheq}
The same procedure can be applied to the matrix model expectation values \eqref{MM:vevs}, which become
\begin{empheq}{equation}\label{MM:vev.def}
	\vev{F(X)}=\frac{1}{N!\,\det I(y,z)}\int\left[\prod_{m=1}^Ndx_m\right]\tilde{\Delta}(x+y)\tilde{\Delta}(x+z)F(gx)\exp\left(-\frac{1}{2}\sum_{n=1}^Nx_n^2\right)\,.
\end{empheq}
We emphazise that the integrals are independent of $y$ and $z$ by virtue of \eqref{MM:Delta.prop}.

It should not come as a surprise that a very convenient choice of polynomials is
\begin{empheq}{equation}
	P_n(x)=\frac{\He_n(x)}{(2\pi)^{\frac{1}{4}}\sqrt{n!}}\,,
\end{empheq}
where $\textrm{He}_n(x)$ are Hermite polynomials.\footnote{Even though we are physicists, we work with the probabilists' version of the Hermite polynomials in order to avoid some awkward factors of $\sqrt{2}$. Some basic properties of the Hermite polynomials are listed in appendix~\ref{Hermite}.} The family $P_n(x)$ is then orthonormal with respect to the Gaussian weight, namely,
\begin{empheq}{equation}
	\int_{-\infty}^{\infty}dx\,P_m(x)P_n(x)\e{-\frac{1}{2}x^2}=\delta_{mn}\,.
\end{empheq}
Furthermore, the matrix elements \eqref{MM:I.def} can be computed explicitly \cite{Gradshteyn}, yielding
\begin{empheq}{equation}\label{MM:I.expl}
	I_{mn}(y,z)
	=\sqrt{\frac{(n-1)!}{(m-1)!}}y^{m-n}\Laguerre^{(m-n)}_{n-1}(-yz)
	=\sqrt{\frac{(m-1)!}{(n-1)!}}z^{n-m}\Laguerre^{(n-m)}_{m-1}(-yz)\,,
\end{empheq}
where $\Laguerre^{(\alpha)}_n(x)$ are Laguerre polynomials. Owing to the properties of these polynomials, the distinction between the cases $m\geq n$ and $m<n$ is not necessary.

\subsection{Properties of the matrix $I(y,z)$}
We now proceed to discuss some noteworthy attributes of the matrix \eqref{MM:I.expl}. First, one easily verifies that 
\begin{empheq}{align}
\label{MM:Iprop.1}
	I(0,0)&=\mathds{1}\,,\\
\label{MM:Iprop.2}
    I^T(y,z)&=I(z,y)\,,\\
\label{MM:Iprop.3}
	I(\xi y,\xi^{-1}z)&=P(\xi) I(y,z)P^{-1}(\xi)~, \qquad P_{mn}(\xi) = \xi^{m} \delta_{mn}\,.
\end{empheq}
Moreover, $I(y,0)$ and $I(0,z)$ are lower and upper triangular matrices, respectively, with unit diagonal entries. Thus, $\det I(y,0)=\det I(0,z)=1$. Some algebra shows that $I(y,z)$ has the LU decomposition
\begin{empheq}{equation}\label{MM:I.LU}
	I(y,z)=I(y,0)I(0,z)\,.
\end{empheq}
It immediately follows that 
\begin{empheq}{equation}\label{MM:I.det}
	\det I(y,z)=1\,.
\end{empheq}

The matrix also satisfies
\begin{empheq}{equation}\label{MM:I.propexp}
	I(y_1,0)I(y_2,0)=I(y_1+y_2,0)\,,
\end{empheq}
which, combined with \eqref{MM:Iprop.2} and \eqref{MM:I.LU}, implies 
\begin{empheq}{equation}
	I(y_1+y_2,z_1+z_2)=I(y_2,0)I(y_1,z_1)I(0,z_2)\,.
\end{empheq}
We also deduce that
\begin{empheq}{equation}\label{MM:I.inverse}
	I^{-1}(y,z)=I(0,-z)I(-y,0)\,.
\end{empheq}
Furthermore, $I(y,z)$ is similar to its inverse, because
\begin{empheq}{equation}\label{MM:I.Similar}
	I^{-1}(y,z)=\left[P(-1)I(0,z)\right]I(y,z)\left[P(-1)I(0,z)\right]^{-1}\,,
\end{empheq}
as can be shown using \eqref{MM:Iprop.3}, \eqref{MM:I.LU} and \eqref{MM:I.inverse}. This implies that the eigenvalues of $I(y,z)$ come in reciprocal pairs and that its characteristic polynomial is either palindromic or anti-palindromic.

The properties \eqref{MM:I.propexp} are reminiscent of an exponential behavior. This is no coincidence. Indeed, one can verify the remarkable relation
\begin{empheq}{equation}\label{MM:I.A}
	I(0,z)=\e{zA}
	\qquad\Rightarrow\qquad
	I(y,z)=\e{yA^T}\e{zA}\,,
\end{empheq}
where $A$ is the matrix given by
\begin{empheq}{equation}
	A_{n,n+1}=\sqrt{n}\,,
\end{empheq}
and all other entries vanishing. More explicitly,
\begin{empheq}{equation}\label{MM:A.expl}
	A=
	\begin{pmatrix}
		0 & \sqrt{1} & 0 & \cdots & 0
		\\
		0 & 0 & \sqrt{2} & \cdots & 0
		\\
		\vdots & \vdots & \vdots & \ddots & \vdots
		\\
		0 & 0 & 0 & \cdots & \sqrt{N-1}
		\\
		0 & 0 & 0 & \cdots & 0
	\end{pmatrix}
	\,.
\end{empheq}
Note that any power of $A$ is traceless and that $A^N=0$. The same is true for $A^T$, of course.

\subsection{Truncated harmonic oscillator}
We promptly notice that $A$ and $A^T$ are nothing more than the matrix representation of the ladder operators of the harmonic oscillator truncated to the first $N$ energy eigenstates. Surely, the number operator
\begin{empheq}{equation}\label{MM:N.def}
	\mathcal{N}=A^T A=
	\begin{pmatrix}
		0 & 0 & 0 & \cdots & 0
		\\
		0 & 1 & 0 & \cdots & 0
		\\
		0 & 0 & 2 & \cdots & 0
		\\
		\vdots & \vdots & \vdots & \ddots & \vdots
		\\
		0 & 0 & 0 & \cdots & N-1
	\end{pmatrix}
\end{empheq}
is diagonal and satisfies
\begin{empheq}{equation}\label{MM:N.A.comm}
	[\mathcal{N},A]=-A\,,
	\qquad
	[\mathcal{N},A^T]=A^T\,.
\end{empheq}
However, $A$ and $A^T$ themselves do not fulfill the Heisenberg algebra. Instead, their commutator reads
\begin{empheq}{equation}
	[A,A^T]=\mathds{1}-N\,\textrm{diag}\left(0,\,\ldots,\,0,\,1\right)\,,
\end{empheq}
which is traceless, as must be for finite-dimensional operators.

The $N$-dimensional quantum mechanical system formed by the operators $A$ and $A^T$, chiefly called the truncated harmonic oscillator, is well known in quantum optics. We recall here some of its properties, following the recent account \cite{Pisanty:2012} and refer the interested reader to more references in that paper. In the next sections we will exploit this connection to the truncated harmonic oscillator in order to extract the leading and sub-leading large $N$ behavior of the circular Wilson loops.

In a fairly obvious notation, we denote by $\ket{n}$, $n=0,\,1,\,\ldots,\,N-1$ the column vector with a $1$ in the $n+1$-th position and zeros elsewhere. These states are eigenstates of the number operator $\mathcal{N}$ and whence are called the number basis. According to \eqref{MM:A.expl}, the ladder operators act upon them as
\begin{empheq}{alignat=7}
\label{MM:A.n.1}
	A\ket{n}&=\sqrt{n}\ket{n-1}
	&\qquad
	n&=0,\,1,\,\ldots,\,N-1\,,
	\\
\label{MM:A.n.2}
	A^T\ket{n}&=\sqrt{n+1}\ket{n+1}
	&\qquad
	n&=0,\,1,\,\ldots,\,N-2\,,
	&\qquad
	A^T\ket{N-1}&=0\,.
\end{empheq}
Of course, the matrix elements \eqref{MM:I.expl} are
\begin{empheq}{alignat=7}
	I_{mn}(y,z)&=\bra{m}\e{yA^T}\e{zA}\ket{n}\,.
\end{empheq}
Actually, this same expression can be obtained by computing the matrix elements of the displacement operator $D(y,z)=e^{ya^{\dagger}}e^{za}=e^{\frac{yz}{2}}e^{ya^{\dagger}+za}$ in the full, infinite-dimensional, quantum harmonic oscillator \cite{Cahill:1969it}, and then truncating to the first $N$ number states.

We will now construct a basis for the $N$-dimensional Hilbert space which is more suited to our purposes. Consider the states
\begin{empheq}{equation} \label{MM:ket.zeta}
	\ket{\zeta}=\sum_{n=0}^{N-1}\frac{\He_n(\zeta)}{\sqrt{n!}}\ket{n}\,,
\end{empheq}
where $\zeta$ is an arbitrary real parameter. By virtue of the recursion relations for the Hermite polynomials, the actions of $A$ and $A^T$ on $\ket{\zeta}$ can be formally represented as
\begin{empheq}{align}
	A\ket{\zeta}&=\left(\zeta-\frac{\partial}{\partial\zeta}\right)\ket{\zeta}-\frac{\He_N(\zeta)}{\sqrt{(N-1)!}}\ket{N-1}\,,\\
	A^T\ket{\zeta}&=\frac{\partial}{\partial\zeta}\ket{\zeta}\,.
\end{empheq}
We see that $\ket{\zeta}$ is an approximate eigenstate of $A+A^T$. It becomes an exact eigenstate if $\zeta$ is a root of $\He_N$,  
\begin{empheq}{equation}\label{MM:A.position}
	\left(A+A^T\right)\ket{\zeta} =\zeta\ket{\zeta}\,,
	\qquad\text{if}\quad
	\He_N(\zeta)=0\,.
\end{empheq}
This construction makes explicit the fact that $A+A^T$ is actually the companion matrix of the Hermite polynomial of degree $N$, namely,
\begin{empheq}{equation}
	\det\left[\zeta -\left(A+A^T\right)\right]=\He_N\left(\zeta\right)\,.
\end{empheq}

Now, the Hermite polynomial $\He_N(\zeta)$ has precisely $N$ distinct roots, which we denote in increasing order by $\zeta_i$, $i=1,\,2,\ldots,\,N$. One can show that the corresponding states $\ket{\zeta_i}$ are linearly independent by computing their inner product with the aid of the Christoffel-Darboux formula,
\begin{empheq}{equation}
	\vev{\eta|\zeta}=\sum_{n=0}^{N-1}\frac{\He_n(\eta)\He_n(\zeta)}{n!}=\frac{1}{(N-1)!}\frac{\He_N(\eta)\He_{N-1}(\zeta)-\He_N(\zeta)\He_{N-1}(\eta)}{\eta-\zeta}\,.
\end{empheq}
Then,
\begin{empheq}{equation}\label{MM:zeta.norm}
	\vev{\zeta_i|\zeta_j}=c_i^2 \delta_{ij}\,, \qquad 
	c_i=\frac{\He_{N+1}(\zeta_i)}{\sqrt{N!}}\,,
\end{empheq}
and the vectors $\ket{\zeta_i}$ form an orthogonal basis of the Hilbert space called the position basis.

The matrix elements of $I(y,z)$ in the position basis are
\begin{empheq}{equation}\label{MM:I.pos}
	I_{ij}(y,z)=\frac1{c_ic_j} \bra{\zeta_i}\e{yA^T}\e{zA}\ket{\zeta_j}\,.
\end{empheq}
We will use this expression extensively in the next sections. Before we move on, we would like to point out that it is possible to write an exact expression for $I^{-1}(y,z)$ that takes a particularly simple form. Indeed, since $A^T$ acts on $|\zeta\rangle$ as a derivative, we have
\begin{empheq}{align}
\notag
	I^{-1}_{ij}(y,z)&=\frac{1}{c_ic_j} \bra{\zeta_i}\e{-zA}\e{-yA^T}\ket{\zeta_j}
	\\
	&=\frac{1}{c_ic_j} \vev{\zeta_i-z|\zeta_j-y}\,.
\end{empheq}
Even though we will not use this formula in what follows, we believe that it could lead to some simplifications in the analysis of sub-leading corrections to the expectation value of the circular Wilson loop.

\subsection{Generating Function}
\label{GF}
The precise relation between the circular Wilson loop and the Gaussian matrix model is
\begin{empheq}{equation}\label{GF:WL.UN}
	\vev{W_R}_{U(N)} =\frac{1}{\textrm{dim}[R]}\vev{\Tr_R \left[\e{X}\right]}\,,
\end{empheq}
where $R$ denotes the representation of the $U(N)$ gauge group and $\Tr_R$ the corresponding trace. The vacuum expectation value on the left is defined on the $\mathcal{N}=4$ SYM theory; the right hand side corresponds to an insertion in the matrix model. 

In this paper we will be concerned with the totally antisymmetric representation of rank $k$ defined by the Young diagram
\begin{empheq}{equation}
	\mathcal{A}_k =\left.\substack{\yng(1,1)\vspace{-0.15cm}\\\vdots\vspace{0.05cm}\\\yng(1)}\right\}k\,.
	\qquad
	k\leq N\,.
\end{empheq}
It has dimension 
\begin{empheq}{equation}
	\textrm{dim}[\mathcal{A}_k]=\binom{N}{k}\,.
\end{empheq}
The structure of the trace $\Tr_{\mathcal{A}_k}$ can get increasingly complicated as the rank $k$ grows. Indeed, for any $N\times N$ matrix $X$ we have
\begin{empheq}{align}
	\Tr_{\mathcal{A}_1} X &=\Tr X\,,
	\\
	\Tr_{\mathcal{A}_2} X &=\frac{1}{2}\left(\Tr X \right)^2-\frac{1}{2}\Tr\left(X^2\right)\,,
	\\
	\Tr_{\mathcal{A}_3} X &=\frac{1}{6}\left(\Tr X\right)^3 -\frac{1}{2}\Tr\left(X^2\right)\Tr X +\frac{1}{3}\textrm{Tr}\left(X^3\right)\,,
\end{empheq}
and so forth. Happily, there is a natural way to encode this structure into the generating function
\begin{empheq}{equation}\label{GF:FA.def}
	F_A(t;X)\equiv\det\left[1+tX\right]=\sum_{k=0}^N t^k \Tr_{\mathcal{A}_k}[X] \,.
\end{empheq}

A straightforward calculation using \eqref{MM:I.def}, \eqref{MM:Deltat.def} and \eqref{MM:vev.def} reveals that the matrix model expectation value of this generating function is 
\begin{empheq}{equation}\label{GF:F.exact}
	\vev{F_A(t;\e{X})}_{U(N)}=\det \left[I(y,z)+t\e{\frac{g^2}{2}}I(g+y,g+z)\right]=\det \left[1+t\e{\frac{g^2}{2}}I(g,g)\right]\,,
\end{empheq}
where the matrix $I(g,g)$ was introduced in \eqref{MM:I.expl}. The dependence on $y$ and $z$ disappears as a consequence of \eqref{MM:I.propexp} and \eqref{MM:I.det}. This is essentially the answer reported by Fiol and Torrents in \cite{Fiol:2013hna}, except that they use a slightly different matrix, which we call $\tilde{I}(g,g)$, obtained from $I(g,g)$ by multiplying the $m$-th row by $\sqrt{(m-1)!}\, g^{-m}$ and dividing the $m$-th column by the same factor. The determinant in \eqref{GF:F.exact} does not change under these operations, so
\begin{empheq}{equation}
	\vev{F_A(t;\e{X})}_{U(N)} = \det\left[ 1 + t \e{\frac12 g^2} \tilde{I}(g,g) \right]~, \qquad 
	\tilde{I}_{mn}(g,g)= \Laguerre^{(m-n)}_{n-1}(-g^2)\,,
\end{empheq}
which is the expression given in \cite{Fiol:2013hna}. Note also that our generating functional differs by a factor of $t^N$ from theirs.

In the answer \eqref{GF:F.exact} we recognize the form of the generating function of antisymmetric traces \eqref{GF:FA.def} for the matrix $I(g,g)$, 
\begin{empheq}{equation}\label{GF:W.trace}
	\vev{F_A(t;\e{X})}=F_A\left(t\e{\frac{\lambda}{8N}};I(g,g)\right)~,
\end{empheq}
which allows us, using \eqref{GF:FA.def} and \eqref{GF:WL.UN}, to obtain the formal yet remarkably simple result
\begin{empheq}{equation}\label{GF:WL.UN.simple}
	\vev{W_{\mathcal{A}_k}}_{U(N)}=\frac{1}{\textrm{dim}[\mathcal{A}_k]} \e{\frac{\lambda k}{8N}}\Tr_{\mathcal{A}_k} [I(g,g)]\,.
\end{empheq}
For the gauge group $SU(N)$, the matrix model must be restricted to traceless matrices. This can be achieved either by using a Lagrange multiplier or by explicitly isolating the trace component in the matrix model integral, as was done in \cite{Gordon:2017dvy}. The outcome is
\begin{empheq}{equation}\label{GF:WL.SUN.simple}
	\vev{W_{\mathcal{A}_k}}_{SU(N)}=\vev{W_{\mathcal{A}_k}}_{U(N)}\e{-\frac{\lambda k^2}{8N^2}}
	=\frac{1}{\textrm{dim}[\mathcal{A}_k]}\e{\frac{\lambda k(N-k)}{8N^2}}\Tr_{\mathcal{A}_k}[I(g,g)]\,.
\end{empheq}
Because the representations $\mathcal{A}_k$ and $\mathcal{A}_{N-k}$ are conjugate to each other for the gauge group $SU(N)$, we expect $\vev{W_{\mathcal{A}_k}}_{SU(N)}=\vev{W_{\mathcal{A}_{N-k}}}_{SU(N)}$. This can be demonstrated by noting that, for an invertible matrix $X$, \eqref{GF:FA.def} implies 
\begin{empheq}{equation}
	\Tr_{\mathcal{A}_{N-k}}[X]=\det[X] \Tr_{\mathcal{A}_k}[X^{-1}]\,.
\end{empheq}
Now, since $I(g,g)$ is similar to its inverse, and $\det I(g,g)=1$, this gives
\begin{empheq}{equation}\label{GF:palin}
	\Tr_{\mathcal{A}_{N-k}}[I(g,g)]= \Tr_{\mathcal{A}_k}[I(g,g)]\,,
\end{empheq}
which proves the assertion. Equation \eqref{GF:palin} also shows that the generating function $F_A(t;I(g,g))$ is a palindromic polynomial. 

The main object of study in the remainder of this paper is the function
\begin{empheq}{equation}\label{GF:F.def}
	\FF(t) = \frac1N \ln F_A(t;I(g,g))= \frac1N \Tr \ln [1+tI(g,g)]\,,
\end{empheq}
from which the traces $\Tr_{\mathcal{A}_k}[I(g,g)]$ can be calculated by
\begin{empheq}{equation}\label{GF:F.traces}
	\Tr_{\mathcal{A}_k}[I(g,g)] = \oint \frac{dt}{2\pi i t} \e{N[\FF(t)-\kappa\ln t]}\,,
\end{empheq}
where we have introduced the ratio
\begin{empheq}{equation}\label{GF:kappa.def}
	\kappa= \frac{k}N~.
\end{empheq}
We are interested in the large $N$ regime, keeping the 't~Hooft coupling $\lambda$ and the ratio $\kappa$ fixed. In this regime, the integral in \eqref{GF:F.traces} is dominated by the saddle point value,\footnote{To obtain \eqref{GF:saddle}, introduce $t = t_\ast \e{iz}$, expand the integrand in $z$ up to second order and evaluate the Gaussian integral. The second term in the exponent comes from the Gaussian integral.} 
\begin{empheq}{equation}
\label{GF:saddle}
	\Tr_{\mathcal{A}_k}[I(g,g)] = \e{N[\mathcal{F}(t_\ast)-\kappa \ln t_\ast ] 
		- \frac12 \ln [ 2\pi N(\kappa + t_\ast^2 \FF''(t_\ast)) ]}~,
\end{empheq}
where $t_\ast$ satisfies the saddle point equation
\begin{empheq}{equation}\label{GF:t.star}
	t_\ast \FF'(t_\ast) = \kappa~. 
\end{empheq} 
Moreover, $\FF(t)$ admits an asymptotic expansion in $1/N$,
\begin{empheq}{equation}\label{GF:F.N.expand}
	\FF(t) = \sum_{n=0}^\infty \FF[n] N^{-n}~.
\end{empheq}
Our aim is to calculate the terms $\FF[0](t)$ and $\FF[1](t)$ from the exact, but formal, expression \eqref{GF:F.def}.

\section{Wilson loops at leading order}
\label{WL0}
In this section, we shall calculate the leading order term $\FF[0](t)$ of the generating function. Of course, $\FF[0](t)$ is known both from a matrix model calculation and from the holographic dual \cite{Yamaguchi:2006tq, Hartnoll:2006is} (the latter implying large $\lambda$ in addition to large $N$). Here, we reproduce it starting from the exact expression \eqref{GF:F.def} by means of two different calculations, which exploit the relation of the matrix $I$ to the truncated harmonic oscillator.

Let us start in the number basis. This calculation will lead to the expression for $\FF[0](t)$ that Okuyama \cite{Okuyama:2017feo} attributes to an unpublished note by Beccaria. 
Starting from \eqref{GF:F.def} and Taylor-expanding the logarithm we may write
\begin{empheq}{equation}\label{WL0:F.powers}
	\FF(t)=-\frac{1}{N}\sum\limits_{n=1}^{\infty} \frac{(-t)^n}{n} \Tr [I(g,g)^n]\,.
\end{empheq}
In order to compute the powers of the matrix $I(g,g)$ we resort to the representation \eqref{MM:I.A}. To leading order in $1/N$, 
commutators of the matrices $A$ and $A^T$ can be neglected, so that we can essentially normal order the product in the trace. This leads to
\begin{empheq}{equation}
\label{WL0:F1}
		\FF(t) = -\sum\limits_{n=1}^\infty \frac{(-t)^n}{n} \frac1N \Tr \left(\e{ngA^T}\e{ngA} \right) +\Order(1/N)\,.
\end{empheq} 
We can now expand the exponentials and calculate the traces. We defer this little calculation to appendix~\ref{calc}.
With the results \eqref{calc:AmAn.tr} and \eqref{calc:AAn.tr2}, this yields
\begin{empheq}{equation}
\label{WL0:F1.N01}
	\FF(t) = -\sum\limits_{n=1}^\infty \frac{(-t)^n}{n} \sum_{k=0}^{N-1} \frac{(ng)^{2k}}{k!(k+1)!} \frac{(N-k)_{k+1}}{N}~.
\end{empheq}
The leading term in $1/N$ is found to be
\begin{empheq}{equation}
\label{WL0:F.Beccaria}
	\FF[0](t) = -\sum_{n=1}^\infty \frac{(-t)^n}{n} \sum_{k=0}^\infty \frac{(n\sqrt{\lambda}/2)^{2k}}{k!(k+1)!} 
	= -\frac2{\sqrt{\lambda}} \sum_{n=1}^\infty \frac{(-t)^n}{n^2} \BesselI[1] (n\sqrt{\lambda})~,
\end{empheq}
where $\BesselI[1](x)$ denotes a modified Bessel function. This is precisely $J_0$ in (2.21) of \cite{Okuyama:2017feo}. 

A slightly simpler way of arriving at \eqref{WL0:F.Beccaria} is to recognize \eqref{WL0:F1} as 
\begin{empheq}{equation}
\label{WL0:F2}
		\FF(t) = -\sum\limits_{n=1}^\infty \frac{(-t)^n}{n} \frac1N \Tr [I(ng,ng)] +\Order(1/N)\,.
\end{empheq} 
The trace follows easily from \eqref{MM:I.expl} and the identity \cite{NIST}
\begin{empheq}{equation}
	\sum_{m=1}^N \Laguerre^{(\alpha)}_{m-1}(x)=\Laguerre_{N-1}^{(\alpha+1)}(x)\,.
\end{empheq}
Hence, \eqref{WL0:F2} becomes
\begin{empheq}{equation}
\label{WL0:F3}
		\FF(t) = -\sum\limits_{n=1}^\infty \frac{(-t)^n}{n} \frac1N \Laguerre^{(1)}_{N-1}(-n^2g^2) +\Order(1/N)\,.
\end{empheq} 
For large order, the Laguerre polynomials satisfy \cite{NIST}
\begin{empheq}{equation} \label{WL0:Lag.limit}
	\lim_{N\to \infty} \frac1{N^\alpha} \Laguerre^{(\alpha)}_N\left(-\frac{z}{N}\right)= \frac1{z^{\frac{\alpha}2}}
	\BesselI[\alpha]\left(2z^\frac12\right)\,,
\end{empheq}
with $\BesselI[\alpha](x)$ being a modified Bessel function. Recalling that $g=\sqrt{\frac{\lambda}{4N}}$, applying the limit \eqref{WL0:Lag.limit} to \eqref{WL0:F3} gives \eqref{WL0:F.Beccaria}.

For a comparison with the matrix model calculation in the saddle point approximation \cite{Hartnoll:2006is}, one can use the following integral representation of the modified Bessel function \cite{NIST},
\begin{equation}
\label{WL0:Bessel.int}
	\BesselI[1] (z) = \frac{z}{\pi} \int\limits_0^\pi \e{z\cos\theta} \sin^2 \theta \rmd \theta~.
\end{equation}
After substituting \eqref{WL0:Bessel.int} into \eqref{WL0:F.Beccaria}, the summation can be carried out to give
\begin{equation}
\label{WL0:F0.int}
	\FF[0] = \frac2{\pi} \int\limits_0^\pi d\theta \sin^2 \theta \ln \left( 1+t \e{\sqrt{\lambda} \cos\theta} \right) ~. 
\end{equation}
A simple change of variables transforms this integral into the one in (2.26) of \cite{Hartnoll:2006is}.

The computation of $\mathcal{F}_0(t)$ turns out to be considerably simpler in the position basis. The reason for this is that, to leading order in $1/N$, we can also write 
\begin{empheq}{equation}
	I(g,g)= \e{g(A+A^T)} + \Order(1/N)\,,
\end{empheq}
where all commutators in the Baker-Campbell-Hausdorff formula contribute to the $\Order(1/N)$ terms. According to \eqref{MM:A.position}, the operator in the exponential is diagonal in the position basis, so that \eqref{GF:F.def} simply becomes
\begin{empheq}{equation}
	\FF(t)=\frac{1}{N}\sum\limits_{i=1}^N\ln\left(1+t\e{g\zeta_i}\right)+ \Order(1/N)\,.
\end{empheq}
The sum is taken over the roots of the Hermite polynomial $\textrm{He}_N(\zeta)$. For large $N$, the distribution of zeros becomes dense and is described by the Wigner semi-circle law. Using formulas \eqref{HeN:roots1} and \eqref{HeN:roots2} to determine the integral measure and dropping the $1/N$ contributions, we get
\begin{empheq}{equation}
	\FF[0](t)=\frac{2}{\pi}\int_0^{\pi}d\theta\,\sin^2\theta\,\ln\left(1+t\e{\sqrt{\lambda}\cos\theta}\right)\,,
\end{empheq}
which agrees with \eqref{WL0:F0.int}.

\section{Wilson loops at next-to-leading order}
\label{WL1}
In this section, we extend the calculation of the generating function $\FF(t)$ to order $1/N$, \ie we obtain the subleading term $\mathcal{F}_1(t)$ in the expansion \eqref{GF:F.N.expand}. Our results agree with \cite{Okuyama:2017feo}, although we also managed to make a slight simplification. In the holographic regime of large $\lambda$, we will find agreement with \cite{Gordon:2017dvy}.

Exploiting the relation between the matrix $I(g,g)$ in the position basis and the zeros of Hermite polynomials appears to be the most promising path to compute the subleading corrections. Starting again from \eqref{WL0:F.powers}, we will proceed as follows. In subsection~\ref{WL1:mat.elements}, we compute the matrix elements of $I(g,g)$ in the position basis up to order $1/N$, which results in an expression like 
\begin{empheq}{equation}
	I(g,g)=I^{(0)}+\frac{1}{N}I^{(1)}+\Order(1/N^2)\,.
\end{empheq}
The leading term $I^{(0)}$ is known from the calculation at the end of the previous section and is diagonal. The main effort in subsection~\ref{WL1:mat.elements} consists in finding $I^{(1)}$. 
We then need to compute powers of $I(g,g)$. This is not as straightforward as one might think because $I^{(1)}$ is a dense matrix. Consider, for example, 
\begin{empheq}{equation}
	I^2(g,g)= [I^{(0)}]^2 +\frac{1}{N}\left[I^{(0)}I^{(1)}+I^{(1)}I^{(0)}\right]
	+\frac{1}{N^2}[I^{(1)}]^2\,.
\end{empheq}
The last term, which naively seems to be of order $1/N^2$, is, in fact, of order $1/N$, because the matrix multiplication contains a sum over $N$ terms of order $1$. The same happens for higher powers. The calculation of the powers of $I(g,g)$ and their traces will be done in subsection~\ref{WL1:powers}. Another source of $1/N$ contributions is the conversion of the sum over the roots of Hermite polynomials, which label the position basis elements, into an integral. We present the details of this conversion in appendix~\ref{SI}. The generating function $\FF(t)$ will be calculated in subsection~\ref{WL1:genfunc}, where we will also provide its integral representation. Finally, in subsecion~\ref{WL1:holo}, we consider $\FF[1](t)$ in the holographic regime of large $\lambda$.

\subsection{Matrix elements}
\label{WL1:mat.elements}

Consider the matrix $I(g,g)=\e{gA^\dagger} \e{gA}$. With $\ket{\zeta}$ defined in \eqref{MM:ket.zeta}, we find by direct application of the properties \eqref{MM:A.n.1} and \eqref{MM:A.n.2}
\begin{equation}
	\e{gA^\dagger} \e{gA} \ket{\zeta} = \sum\limits_{k=0}^{N-1}\sum\limits_{l=0}^{N-1} 
	\sum\limits_{n=0}^{\min(N-1, N-1-l+k)} \frac{g^{k+l}}{k! l!} 
	\binom{n}{k} k! \frac{\He_{n+l-k}(\zeta)}{\sqrt{n!}} \ket{n}~.
\end{equation}
Reordering the summations such that the sum over states stays on the left, one obtains
\begin{equation}
\label{WL1:klsum.todo}
	\e{gA^\dagger} \e{gA} \ket{\zeta} = \sum\limits_{n=0}^{N-1}\sum\limits_{k=0}^{n} 
	\sum\limits_{l=0}^{N-1-n+k} \frac{g^{k+l}}{l!} 
	\binom{n}{k} \frac{\He_{n+l-k}(\zeta)}{\sqrt{n!}} \ket{n}~.
\end{equation}
\begin{figure}[ht]
	\begin{center}
		\includegraphics[width =0.8\textwidth]{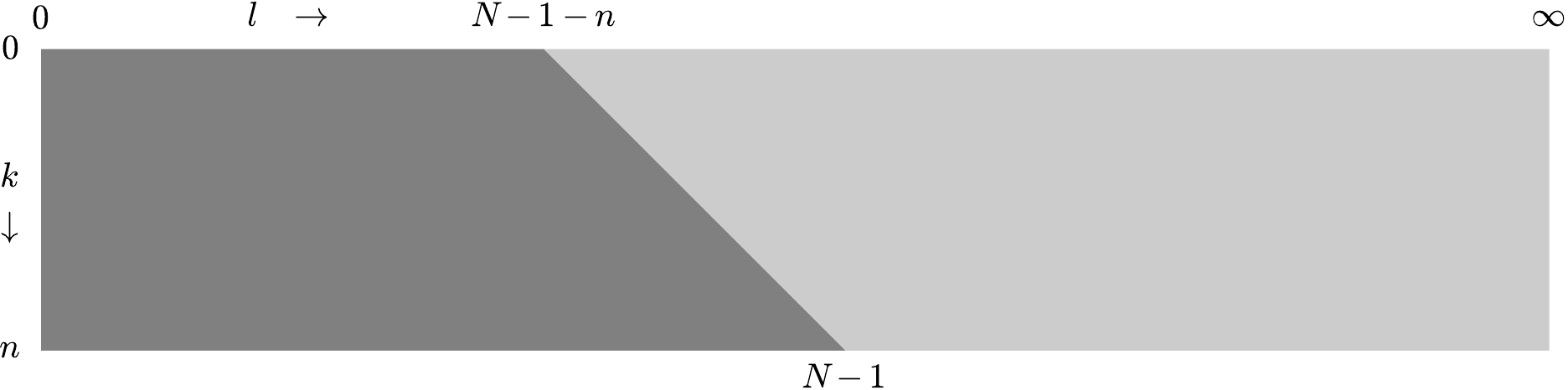}
	\end{center}
	\caption{Illustration of the domain of summations over $k$ and $l$ in \eqref{WL1:klsum.todo}. The domain to be summed over is the dark shaded area. Instead, we sum over the semi-infinite rectangle and subtract the sum over the light shaded area.
	\label{WL1:kl.fig}}
\end{figure}
Note that the binomial coefficient restricts the sum over $k$. The next step consists in evaluating the sums over $k$ and $l$. These sums, in the $k,l$ plane, extend over a domain that can be represented by the dark shaded trapezoid area in Fig.~\ref{WL1:kl.fig}. The trick is to extend the summation domain to a semi-infinite stripe until $l=\infty$ and subtract the sum over the domain illustrated by the light shaded area. In the sum over the semi-infinite stripe, it comes handy to introduce $r=k+l$ and to sum over $r$ instead of $l$. This yields
\begin{equation}
	\sum\limits_{k=0}^{n} \sum\limits_{l=0}^{\infty} \frac{g^{k+l}}{l!}
	\binom{n}{k} \He_{n+l-k}(\zeta) = 
	\sum\limits_{r=0}^{\infty} \sum\limits_{k=0}^{n} \frac{g^{r}}{r!} \binom{r}{k}
	\binom{n}{k} k! \He_{n+r-2k}(\zeta)~.
\end{equation}
Here, we recognize the linearization formula \eqref{HeN:lin.form} and, subsequently, the generating function for the Hermite polynomials \eqref{HeN:gen.func}. Therefore, the sum over the semi-infinite stripe yields the simple expression
\begin{equation}
\label{WL1:sum.stripe}
	\sum\limits_{k=0}^{n} \sum\limits_{l=0}^{\infty} \frac{g^{k+l}}{l!}
	\binom{n}{k} \He_{n+l-k}(\zeta) = \e{g\zeta-\frac12 g^2} \He_n(\zeta)~.
\end{equation}

In the sum that must be subtracted (the light shaded area), we introduce the variable $s=l-k+n-N$ and get
\begin{equation}
\label{WL1:sum.subtract}
	\sum\limits_{k=0}^{n} \sum\limits_{l=N-n+k}^{\infty} \frac{g^{k+l}}{l!}
	\binom{n}{k} \He_{n+l-k}(\zeta) = 
	\sum\limits_{s=0}^{\infty} \sum\limits_{k=0}^{n} \frac{g^{N-n+s+2k}}{(N-n+s+k)!} \binom{n}{k}
	\He_{N+s}(\zeta)~.
\end{equation}
Notice that, until now, it was not necessary to require that $\zeta$ be a root of $\He_N$. From now on, we will assume that it is. This implies that the $s=0$ term in the sum in \eqref{WL1:sum.subtract} vanishes. Furthermore, it allows us to express the Hermite polynomials $\He_{N+s}(\zeta)$ as a sum of the polynomials $\He_n(\zeta)$ with $n<N$. This can be shown by means of the linearization formula \eqref{HeN:lin.form}, setting $n=N$. More precisely, we can write 
\begin{equation}
\label{WL1:He.Ns}
	\He_{N+s}(\zeta) = - \sum\limits_{l=0}^{[\frac{s-1}2]} a_{s,l} \He_{N-s+2l}(\zeta)~, 
	\qquad \text{for} \quad \He_N(\zeta)=0\,,
\end{equation}
with coefficients $a_{s,l}$, which satisfy
\begin{equation}
\label{WL1:a.prop}
	\sum\limits_{k=0}^m \binom{s}{k} \binom{N}{k} k! a_{s-2k, m-k} = \binom{s}{m} \binom{N}{s-m} (s-m)!~.
\end{equation}
The coefficients $a_{s,l}$ can be determined applying \eqref{WL1:a.prop} recursively for increasing $m$. The first steps of this recursion yield
\begin{equation}
\label{WL1:a.rec}
	a_{s,0} = \binom{N}{s}s!~,\qquad a_{s+2,1} = - \binom{N}{s}s! s(s+2)~,\qquad  \ldots~.
\end{equation}
We shall not investigate this recursion further, because it will turn out that only $a_{s,0}$ is relevant for our purposes.
After substituting \eqref{WL1:He.Ns} into \eqref{WL1:sum.subtract}, we can reorder the summations over $s$ and $l$ by introducing $r=s-2l$ and obtain
\begin{multline}
\label{WL1:sum.subtract2}
	- \sum\limits_{s=0}^{\infty} \sum\limits_{k=0}^{n} \frac{g^{N-n+s+2k}}{(N-n+s+k)!} \binom{n}{k}
	\sum\limits_{l=0}^{[\frac{s-1}2]} a_{s,l} \He_{N-s+2l}(\zeta) = \\
	- \sum\limits_{k=0}^{n} \sum\limits_{r=1}^{N} \binom{n}{k} \He_{N-r}(\zeta) \sum\limits_{l=0}^\infty 
	\frac{g^{N-n+r+2l+2k}}{(r+N-n+2l+k)!} a_{r+2l,l}~.
\end{multline}

With these results, let us now consider the matrix element of $\e{gA^\dagger}\e{gA}$ between two (non-normalized) position eigenstates. Remember that the sum over $k$ and $l$ in \eqref{WL1:klsum.todo} is given by \eqref{WL1:sum.stripe} minus \eqref{WL1:sum.subtract2}. We find
\begin{align}
\label{WL1:mat.el1}
	\bra{\zeta_i} \e{gA^\dagger}\e{gA}\ket{\zeta_j} &= \e{g\zeta_j-\frac12 g^2} \braket{\zeta_i}{\zeta_j} \\
\notag
	&\quad + \sum\limits_{r,s=1}^N \He_{N-s}(\zeta_i) \He_{N-r}(\zeta_j) 
	\sum\limits_{k=0}^\infty \sum\limits_{l=0}^\infty \frac{g^{r+s+2(l+k)} a_{r+2l,l}}{k!(N-s-k)!(s+r+2l+k)!}~. 
\end{align}
where we have used the definition of the states $\ket{\zeta}$ \eqref{MM:ket.zeta}, extended the summation over $k$ (this can be done because of the binomial) and let $s=N-n$. Furthermore, rewriting the sums over $l$ and $k$ as a sum over $n=l+k$ and $k$  gives
\begin{align}
\label{WL1:mat.el2}
	\bra{\zeta_i} \e{gA^\dagger}\e{gA}\ket{\zeta_j} &= \e{g\zeta_j-\frac12 g^2} \braket{\zeta_i}{\zeta_j} \\
\notag
	&\quad + \sum\limits_{r,s=1}^N  \He_{N-s}(\zeta_i) \He_{N-r}(\zeta_j) 
	\sum\limits_{n=0}^\infty \frac{g^{r+s+2n}}{(r+s+2n)!} 
	\sum\limits_{k=0}^n \binom{r+s+2n}{k} \frac{a_{r+2(n-k),n-k}}{(N-s-k)!}~. 
\end{align}
In the first term on the right hand side, one can recognize the leading order result, corrected by the exponential $\e{-\frac12 g^2}$. Therefore, the second term is of order $1/N$, so that leading order relations can be used to manipulate it. 
A short inspection of the sum over $k$ in \eqref{WL1:mat.el2} for some low values of $n$ shows that the leading order contribution in $1/N$ comes from the term with $k=n$. Therefore, using \eqref{WL1:a.rec}, we get
\begin{align}
\notag
	  \sum\limits_{k=0}^n \binom{r+s+2n}{k} \frac{a_{r+2(n-k),n-k}}{(N-s-k)!} 
	  &\sim \binom{r+s+2n}{n} \frac{N!(N-s-n+1)_n}{(N-r)!(N-s)!} \\
\label{WL1:sum.k}
	  &\sim \binom{r+s+2n}{n} \frac{N!N^n}{(N-r)!(N-s)!}~, 
\end{align}
where subleading terms in $1/N$ have been omitted.  Then, after normalizing the matrix elements \eqref{WL1:mat.el2} by means of \eqref{MM:zeta.norm} and substituting \eqref{WL1:sum.k}, we obtain
\begin{align}
\label{WL1:mat.el3}
	I_{ij}(g,g) &= \e{g\zeta_j-\frac12 g^2} \delta_{ij} \\
\notag
	&\quad + \sum\limits_{r,s=1}^N \frac{(N!)^2}{(N-r)!(N-s)!} 
	\frac{\He_{N-s}(\zeta_i) \He_{N-r}(\zeta_j)}{\He_{N+1}(\zeta_i) \He_{N+1}(\zeta_j)}
	\sum\limits_{n=0}^\infty \frac{g^{r+s+2n} N^n}{(r+s+n)!n!} +\Order\left(\frac1{N^2}\right)~. 
\end{align}

It is possible to obtain simple expressions for the ratios $\He_{N-s}(\zeta_i)/ \He_{N+1}(\zeta_i)$. We include the calculation in Appendix~\ref{HeNU}, where we show that, to leading order in $1/N$, 
\begin{equation}
\label{WL1:Hermite.Chebychev}
	\frac{\He_{N-s}(\zeta)}{\He_{N+1}(\zeta)} \sim - \frac{(N-s)!}{N!} N^{\frac{s-1}2} \ChebU_{s-1}(\cos \theta)~,
\end{equation}
where $\cos\theta$ is determined by $\zeta \sim 2 \sqrt{N}\cos \theta$, and $\ChebU_s$ denote the Chebychev polynomials of the second kind \cite{Gradshteyn},
\begin{equation}
\label{WL1:Chebychev}
	\ChebU_s(\cos \theta) = \frac{\sin[(s+1)\theta]}{\sin\theta}~.
\end{equation}
Hence, we obtain our final result for the matrix $I(g,g)$ in the position basis,
\begin{align}
\notag
	I_{ij}(g,g) &= \e{g\zeta_j-\frac12 g^2} \delta_{ij} 
 + \frac1N \sum\limits_{r,s=1}^N  \frac{\sin(s\theta_i) \sin(r\theta_j)}{\sin \theta_i \sin \theta_j} 
	\sum\limits_{n=0}^\infty \frac{(\sqrt{\lambda}/2)^{r+s+2n}}{(r+s+n)!n!} + \Order\left(\frac1{N^2}\right) \\
\label{WL1:mat.el4}
	&= \e{-\frac{\lambda}{8N}} \e{\sqrt{\lambda}\sqrt{1+\frac1{2N}} \cos \theta_i} \delta_{ij} 
	+ \frac1N \sum\limits_{r,s=1}^\infty  \frac{\sin(s\theta_i) \sin(r\theta_j)}{\sin \theta_i \sin \theta_j} 
	\BesselI[r+s] (\sqrt{\lambda}) + \Order\left(\frac1{N^2}\right)~.
\end{align}
In the step from the first to the second line we have recognized the modified Bessel functions in the sums over $n$ and expressed the roots $\zeta_i$ in the diagonal term in terms of $\cos\theta_i$ being careful to use the expression \eqref{HeN:roots1}, which is exact to order $1/N$,\footnote{In the second term, which is already of order $1/N$, this care was not needed.}. Moreover, we have extended the summations to infinity in the sense of an asymptotic expansion.

\subsection{Powers of $I(g,g)$ and their traces}
\label{WL1:powers}

In this subsection, we return to calculate the quantities needed in the generating function \eqref{WL0:F.powers}, namely the powers of the matrix $I(g,g)$ and their traces. As mentioned at the beginning of this section, calculating the powers of $I(g,g)$ to order $1/N$ is not straightforward. Fortunately, after calculating a few powers, \eg $I^2(g,g)$ and $I^3(g,g)$, one recognizes a pattern that can then be proven by induction for all powers. The calculation is somewhat tedious and involves the facts that the modified Bessel functions have the generating function
\begin{equation}
\label{WL1:Bessel.gen}
	\e{x\cos\theta} = \sum\limits_{k=-\infty}^\infty \BesselI[k](x) \cos(k\theta)~, 
\end{equation}
and that they satisfy the addition theorem \cite{NIST}
\begin{equation}
\label{WL1:Bessel.summation.theorem}
	\BesselI[n](x+y) = \sum\limits_{k=-\infty}^\infty \BesselI[n-k](x) \BesselI[k](y)~.
\end{equation}
Dropping $1/N^2$ terms, the result is summarized in the formula
\begin{align}
\label{WL1:mat.Mm}
	I^m_{ij}(g,g) 
	&= \e{-\frac{m\lambda}{8N}} \e{m\sqrt{\lambda}\sqrt{1+\frac1{2N}} \cos \theta_i} \delta_{ij} 
	+ \frac1N \sum\limits_{r,s=1}^\infty  \frac{\sin(r\theta_i) \sin(s\theta_j)}{\sin \theta_i \sin \theta_j} 
	\left[ \BesselI[r+s] (m\sqrt{\lambda}) + M^{(m)}_{rs} \right]~.
\end{align}
Here, $M^{(m)}_{rs}$ denote some (infinite) matrices, which satisfy the recursion relations  
\begin{equation}
\label{WL1:M.rec}
	M^{(m+1)}_{rs} = \sum\limits_{t=1}^\infty M^{(m)}_{rt} \BesselI[s-t](\sqrt{\lambda}) 
		- \sum\limits_{t=0}^\infty \BesselI[r+t](m\sqrt{\lambda}) \BesselI[s+t](\sqrt{\lambda})
\end{equation}
and $M^{(1)}_{rs}=0$.

Let us evaluate the trace of \eqref{WL1:mat.Mm}. This calculation is nearly identical to the one resulting in \eqref{WL1:trace.check}, in particular with respect to the cancellation of various $1/N$ contributions, and yields
\begin{equation}
\label{WL1:Mm.trace}
	\frac1N \Tr I^m(g,g) =  \frac2{m\sqrt{\lambda}} \BesselI[1](m\sqrt{\lambda}) \e{-\frac{m\lambda}{8N}} 
	+ \frac1N \sum\limits_{r=1}^\infty M^{(m)}_{rr}~.
\end{equation}
The first term on the right hand side reproduces the leading order expression \eqref{WL0:F.Beccaria}, with an exponential correction that can be traced back to the exponential in \eqref{GF:W.trace}, which is missing in \eqref{GF:F.def}. 
The second term, which is the substantial part of the $1/N$ contributions, will be evaluated in the remainder of this subsection.

Henceforth, let $z=\sqrt{\lambda}$ in order to simplify the notation. For $m=2$, we simply have from \eqref{WL1:M.rec}
\begin{equation}
\notag 
	\sum\limits_{r=1}^\infty M^{(2)}_{rr} = 
		- \sum\limits_{r=1}^\infty \sum\limits_{t=0}^\infty \BesselI[r+t](z) \BesselI[r+t](z)\\
	= - \sum\limits_{v=1}^\infty \sum\limits_{r=1}^v \BesselI[v](z) \BesselI[v](z)~,
\end{equation}  
where we have re-ordered the summations over $r$ and $t$. For $m>2$, using the same reordering,
\eqref{WL1:M.rec} leads to the following pattern,
\begin{equation}
\label{WL1:M.pattern}
	\sum\limits_{r=1}^\infty M^{(m)}_{rr} = - \sum\limits_{a=1}^{m-1} \sum\limits_{v=1}^\infty 
		\BesselI[v]((m-a)z) S_a(v;z)~,
\end{equation}
where $S_a(v;z)$ stands for 
\begin{equation}
\label{WL1:S1} 
	S_1(v;z) = v \BesselI[v](z)
\end{equation}
and
\begin{equation}
\label{WL1:M.sums}
	S_a(v;z) = \sum\limits_{r=1}^v \sum\limits_{t_1=1}^\infty  \cdots \sum\limits_{t_{a-1}=1}^\infty 
		 	\BesselI[v-r+t_1](z) \BesselI[t_2-t_1](z) \cdots \BesselI[r-t_{a-1}](z)\,,
		 	\qquad a = 2,3\ldots, m-1\,.
\end{equation}

Let us calculate \eqref{WL1:M.sums} for $a=2$, where there is only one $t$-summation. The simplest way to proceed is to use the invariance of the summand under the transformation $r\to v-r+1$, $t\to 1-t$. This yields
\begin{equation}
\label{WL1:S2}
	S_2(v;z) = \frac12 \sum\limits_{r=1}^v \sum\limits_{t=-\infty}^\infty \BesselI[v-r+t](z) \BesselI[t-r](z)
	= \frac12 v \BesselI[v](2z)~, 
\end{equation}
where we have recognized the summation formula \eqref{WL1:Bessel.summation.theorem}.

Unfortunately, the same trick does not suffice to easily obtain the nested sums for $a>2$. However, using a generating function for $S_a(v;z)$, we prove in appendix~\ref{SA} that
\begin{equation}
\label{WL1:Sa}
 	S_a(v;z) = \frac{v}{a} \BesselI[v](az)~.
\end{equation}
This is a remarkable result, which we have not found in the literature. With \eqref{WL1:Sa}, we can return to \eqref{WL1:M.pattern}, which simplifies to
\begin{align}
\notag
	\sum\limits_{r=1}^\infty M^{(m)}_{rr} &= - \sum\limits_{a=1}^{m-1} \sum\limits_{v=1}^\infty 
		\frac{v}{a} \BesselI[v]((m-a)z) \BesselI[v](az) \\
\notag 
	&= -\frac{z}{2} \sum\limits_{a=1}^{m-1} \sum\limits_{v=1}^\infty \BesselI[v]((m-a)z) 
	\left[ \BesselI[v-1](az) - \BesselI[v+1](az)\right] \\
\notag
	&= -\frac{z}{2} \sum\limits_{a=1}^{m-1} \sum\limits_{v=1}^\infty \left[ \BesselI[v]((m-a)z) 
	\BesselI[v-1](az) - \BesselI[v+1]((m-a)z) \BesselI[v](az) \right] \\
\label{WL1:Mm.tracem}
	& = - \frac{z}2 \sum\limits_{a=1}^{m-1} \BesselI[0](az) \BesselI[1][(m-a)z]~.
\end{align}

\subsection{Generating function}
\label{WL1:genfunc}

Putting together \eqref{WL0:F.powers}, \eqref{WL1:Mm.trace} and \eqref{WL1:Mm.tracem}, we obtain the generating function $\FF(t)$ to order $1/N$,
\begin{equation}
\label{WL1:F1}
	\FF(t)	= - \sum_{m=1}^\infty (-t)^m \left[ \frac2{m^2 \sqrt{\lambda}} \BesselI[1](m\sqrt{\lambda}) \e{-\frac{m\lambda}{8N}} 
	- \frac1N \frac{\sqrt{\lambda}}{2m} \sum\limits_{a=1}^{m-1} \BesselI[0](a\sqrt{\lambda})\BesselI[1]((m-a)\sqrt{\lambda}) 
	\right] + \Order(1/N^2)~.
\end{equation}
Let us compare \eqref{WL1:F1} with Okuyama's result, which is (2.21) of \cite{Okuyama:2017feo}.
The first term in the brackets reproduces $J_0$ of \cite{Okuyama:2017feo}, except for the exponential factor, which arises from the fact the our $\FF(t)$ is not exactly the generating function of $U(N)$ Wilson loops, but is defined in terms of the palindromic polynomial $F_A(t;I(g,g))$, c.f.~\eqref{GF:F.def}. The second term in the brackets can easily be shown to reproduce $J_1$ of \cite{Okuyama:2017feo}, because
\begin{align}
\notag 
	&\frac{\partial}{\partial \lambda} \sum\limits_{a=1}^{m-1} 
	\left[ \sqrt{\lambda} \BesselI[0](a\sqrt{\lambda})\BesselI[1]((m-a)\sqrt{\lambda})\right]
	 \\
\notag
	&= \frac12 \sum\limits_{a=1}^{m-1} \left[ a \BesselI[1](a\sqrt{\lambda})\BesselI[1]((m-a)\sqrt{\lambda}) + (m-a) \BesselI[0](a\sqrt{\lambda})\BesselI[0]((m-a)\sqrt{\lambda})\right] \\
	&= \frac{m}4 \sum\limits_{a=1}^{m-1} \left[ \BesselI[1](a\sqrt{\lambda})\BesselI[1]((m-a)\sqrt{\lambda}) + \BesselI[0](a\sqrt{\lambda})\BesselI[0]((m-a)\sqrt{\lambda})\right]~,
\end{align}
which appears in the integrand in $J_1$. The step from the second to the third line consists in symmetrizing the summands with respect to $a\to m-a$. We note that our result for the $1/N$ term is slightly simpler than Okuyama's, because it does not involve an integral.

In the remainder of this section, we will find an integral representation of $\FF[1](t)$.\footnote{The integral representation of $\FF[0](t)$ is \eqref{WL0:F0.int}.}
Consider the first term in brackets in \eqref{WL1:F1}. Because it contains $\FF[0](t)$, but also corrections in $1/N$, we will denote it by $\widetilde{\FF[0]}(t)$. After using the integral representation of the modified Bessel function \eqref{WL0:Bessel.int}, the sum over $m$ can be performed, which yields
\begin{align}
\notag 
	\widetilde{\FF[0]}(t) &= \frac2{\pi} \int\limits_0^\pi \rmd \theta \sin^2 \theta \ln 
		\left( 1 + t \e{\sqrt{\lambda} \cos \theta - \frac{\lambda}{8N}} \right) \\
\label{WL1:tFF0.integ}
	&= \FF[0] - \frac{\lambda}{4\pi N} \int\limits_0^\pi \rmd \theta \sin^2 \theta \frac{t\e{\sqrt{\lambda}\cos\theta}}{%
	1+t \e{\sqrt{\lambda}\cos\theta}}
	+ \Order(1/N^2)~.
\end{align}

Now, consider the second term in brackets in \eqref{WL1:F1}, which we shall denote by $\widetilde{\FF[1]}(t)/N$. Using the standard integral representation for $\BesselI[0]$ and $\BesselI[1]$ \cite{NIST}, the sum over $a$ can be done, which gives 
\begin{equation}
\label{WL1:tFF1.2}
	\widetilde{\FF[1]}(t) = \frac{\sqrt{\lambda}}{2\pi^2} \sum\limits_{m=1}^\infty \frac{(-t)^m}{m} 
	\int\limits_0^\pi \rmd \theta \int\limits_0^\pi \rmd \phi \cos \phi \,
	\frac{\e{\sqrt{\lambda}(\cos\theta+m\cos\phi)} - \e{m\sqrt{\lambda}(m\cos\theta+\cos\phi)}}{%
	\e{\sqrt{\lambda}\cos\phi}-\e{\sqrt{\lambda}\cos\theta}}~. 
\end{equation}
Lets us rewrite the fraction in the integrand as
\begin{equation}
\notag
	\frac{\e{\sqrt{\lambda}(\cos\theta+m\cos\phi)} - \e{m\sqrt{\lambda}(m\cos\theta+\cos\phi)}}{%
	\e{\sqrt{\lambda}\cos\phi}-\e{\sqrt{\lambda}\cos\theta}}
	= 
	\frac{\e{\sqrt{\lambda}(\cos\theta-\cos\phi)}}{%
	1-\e{\sqrt{\lambda}(\cos\theta-\cos\phi)}} 
	\left(\e{m\sqrt{\lambda}\cos\phi}- \e{m\sqrt{\lambda}\cos\theta} \right) -  \e{m\sqrt{\lambda}\cos\theta} 
\end{equation}
The last term on the right hand side, which does not depend on $\phi$, integrates to zero in the $\phi$-integral. For the remaining term, performing the sum over $m$ in \eqref{WL1:tFF1.2} leads to 
\begin{equation}
\label{WL1:tFF1.3}
	\widetilde{\FF[1]}(t) = - \frac{\sqrt{\lambda}}{2\pi^2} 
	\int\limits_0^\pi \rmd \theta \int\limits_0^\pi \rmd \phi \cos \phi \,
	\frac{\e{\sqrt{\lambda}(\cos\theta-\cos\phi)}}{1-\e{\sqrt{\lambda}(\cos\theta-\cos\phi)}}
	\ln \frac{1 + t \e{\sqrt{\lambda} \cos \phi}}{1 + t \e{\sqrt{\lambda} \cos \theta}}~. 
\end{equation}
Thus, combining \eqref{WL1:tFF0.integ} with \eqref{WL1:tFF1.3}, we obtain the integral representation for $\FF[1](t)$ as 
\begin{align}
\label{WL1:FF1.integ}
	\FF[1](t) &= - \frac{\sqrt{\lambda}}{2\pi^2} 
	\int\limits_0^\pi \rmd \theta \int\limits_0^\pi \rmd \phi \cos \phi 
	\frac{\e{\sqrt{\lambda}(\cos\theta-\cos\phi)}}{1-\e{\sqrt{\lambda}(\cos\theta-\cos\phi)}}
	\ln \frac{1 + t \e{\sqrt{\lambda} \cos \phi}}{1 + t \e{\sqrt{\lambda} \cos \theta}}\\
\notag
	& \quad - \frac{\lambda}{4\pi} \int\limits_0^\pi \rmd \theta \sin^2 \theta \frac{t\e{\sqrt{\lambda}\cos\theta}}{%
	1+t \e{\sqrt{\lambda}\cos\theta}}~. 
\end{align}

\subsection{Holographic regime}
\label{WL1:holo}

Our aim in this subsection is to evaluate the saddle point value of \eqref{WL1:FF1.integ} in the regime of large $\lambda$. We remind the reader that the saddle point, in this regime, is given by 
\begin{equation}
\label{WL1:t.saddle}
	t =\e{-\sqrt{\lambda}\cos \ths}~,
\end{equation}
where the angle $\ths$ satisfies 
\begin{equation}
\label{WL1:kappa}
	\kappa \equiv \frac{k}{N} = \frac1{\pi} \left( \ths - \frac12 \sin 2\ths \right)~.
\end{equation}
Let us start with integral on the second line of \eqref{WL1:FF1.integ}. Dropping the terms that are exponentially suppressed for large $\lambda$, we easily obtain
\begin{equation}
\label{WL1:tFF0.holo}
	- \frac{\lambda}{4\pi} \int\limits_0^\pi \rmd \theta \sin^2 \theta \frac{t\e{\sqrt{\lambda}\cos\theta}}{%
	1+t \e{\sqrt{\lambda}\cos\theta}} = 
	- \frac{\lambda}{4\pi} \int\limits_0^{\ths} \rmd \theta \sin^2 \theta =  - \frac{\lambda \kappa}{8}~.
\end{equation}
\begin{figure}
\begin{center}
\includegraphics[width=0.4\textwidth]{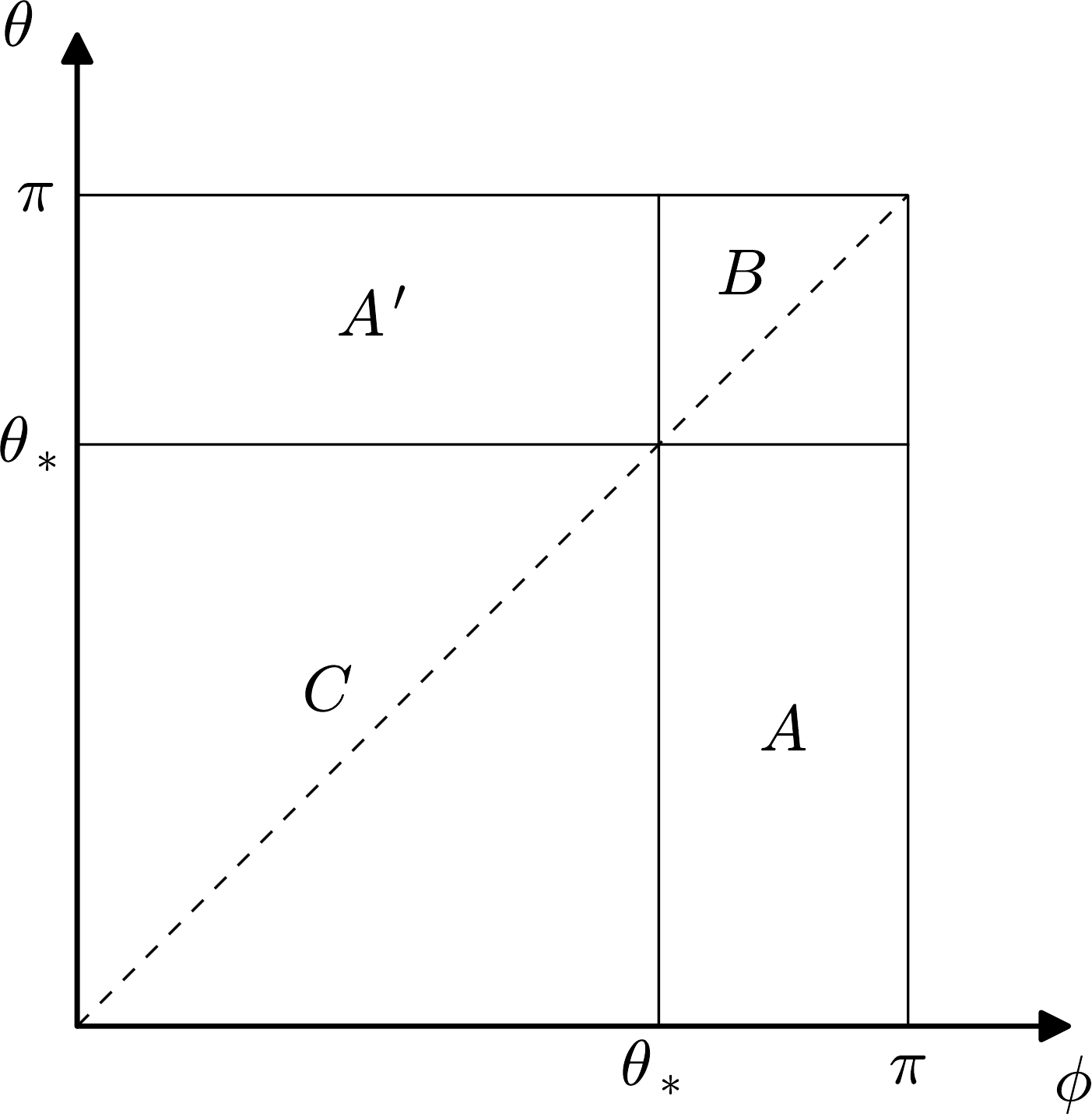}
\caption{Division of the integration domain for the double integral in \eqref{WL1:FF1.integ}. On the dashed line, where $\theta=\phi$, both, the denominator and the logarithm in the integral vanish.\label{WL1:plane.fig}}
\end{center}
\end{figure}
To evaluate the double integral on the first line of \eqref{WL1:FF1.integ}, we divide the integration domain into the four regions $A$, $A'$, $B$ and $C$ illustrated in Fig.~\ref{WL1:plane.fig}. What differs between these four regions is the approximation of the logarithm in the integrand. In the region $A$ ($\theta<\ths<\phi$), we have
\begin{equation}
\label{WL1:ln.A}
	A:\quad \ln \frac{1 + \e{\sqrt{\lambda}(\cos\phi-\cos \ths)}}{1 + \e{\sqrt{\lambda}(\cos\theta-\cos\ths)}} 
	\approx - \sqrt{\lambda} (\cos\theta-\cos\ths)~,
\end{equation} 
while the fraction in front of the logarithm is approximately $-1$. Therefore, the contribution of the region $A$ is 
\begin{equation}
\label{WL1:integ.A}
	A = - \frac{\lambda}{2\pi^2} \int\limits_0^{\ths} \rmd \theta \int\limits_{\ths}^\pi \rmd \phi 
	\cos \phi\left( \cos\theta - \cos \ths\right)  
	= \frac{\lambda}{2\pi^2}  \left( \sin^2 \ths - \frac12 \ths \sin 2\ths \right)~.	
\end{equation}
In the region $A'$ ($\phi<\ths<\theta$), we have
\begin{equation}
\label{WL1:ln.Ap}
	A':\quad \ln \frac{1 + \e{\sqrt{\lambda}(\cos\phi-\cos \ths)}}{1 + \e{\sqrt{\lambda}(\cos\theta-\cos\ths)}} 
	\approx \sqrt{\lambda} (\cos\phi-\cos\ths)~,
\end{equation} 
but the fraction in front of the logarithm is exponentially suppressed. Hence,
\begin{equation}
\label{WL1:integ.Ap}
	A'=0~.
\end{equation}
Next, look at region $B$, where $\theta>\ths$ and $\phi>\ths$. Here,
\begin{equation}
\label{WL1:ln.B}
	B:\quad \ln \frac{1 + \e{\sqrt{\lambda}(\cos\phi-\cos \ths)}}{1 + \e{\sqrt{\lambda}(\cos\theta-\cos\ths)}} 
	\approx \e{\sqrt{\lambda} (\cos\phi-\cos\ths)} \left( 1- \e{\sqrt{\lambda}(\cos \theta-\cos \phi)}\right)~.
\end{equation}
The term in parentheses precisely cancels the denominator of the term in front of the logarithm, and what remains of the integrand is again exponentially suppressed for large $\lambda$. Thus,
\begin{equation}
\label{WL1:integ.B}
	B=0~.
\end{equation}
In the remaining region $C$, where $\theta<\ths$ and $\phi<\ths$, we have
\begin{equation}
\label{WL1:ln.C}
	C:\quad \ln \frac{1 + \e{\sqrt{\lambda}(\cos\phi-\cos \ths)}}{1 + \e{\sqrt{\lambda}(\cos\theta-\cos\ths)}} 
	\approx \sqrt{\lambda} \left(\cos\phi-\cos\theta\right)~.
\end{equation}
Therefore, we can write the contribution from the region $C$ to the integral as  
\begin{equation}
\label{WL1:integ.C.1}
	C = -\frac{\lambda}{2\pi^2} \int\limits_0^{\ths} \rmd \theta \int\limits_0^{\ths} \rmd \phi 
	\cos \phi\left( \cos\phi - \cos \theta \right) 
	\frac{\e{\sqrt{\lambda}\cos\theta}}{\e{\sqrt{\lambda}\cos\phi}-\e{\sqrt{\lambda}\cos\theta}}~.
\end{equation}
Because the integration domain in \eqref{WL1:integ.C.1} is symmetric with respect to $\phi$ and $\theta$, we can symmetrize the integrand and rewrite \eqref{WL1:integ.C.1} as
\begin{equation}
\label{WL1:integ.C.2}
	C = \frac{\lambda}{8\pi^2} \int\limits_0^{\ths} \rmd \theta \int\limits_0^{\ths} \rmd \phi 
	\left[ (\cos\phi - \cos \theta)^2 + (\cos^2 \theta -\cos^2 \phi)  
	\frac{\e{\sqrt{\lambda}\cos\phi} +\e{\sqrt{\lambda}\cos\theta}}{\e{\sqrt{\lambda}\cos\phi}-\e{\sqrt{\lambda}\cos\theta}}
	\right]~.
\end{equation}
The first term is readily integrated and yields
\begin{equation}
\label{WL1:integ.C1}
	C_1 = \frac{\lambda}{8\pi^2} \int\limits_0^{\ths} \rmd \theta \int\limits_0^{\ths} \rmd \phi \,
	(\cos\phi - \cos \theta)^2 =
	\frac{\lambda}{8\pi^2} \left( \ths^2 +\frac12 \ths \sin 2 \ths - 2 \sin^2 \ths \right)~.
\end{equation}
In the second term in \eqref{WL1:integ.C.2}, we write $\cos^2 \theta -\cos^2 \phi= \sin^2 \phi -\sin^2\theta$ and realize that, using the symmetry of the integrand and the integration domain, we can write 
\begin{equation}
\label{WL1:integ.C2.1}
	C_2 = \frac{\lambda}{4\pi^2} \int\limits_0^{\ths} \rmd \theta \int\limits_0^{\ths} \rmd \phi 
	\sin^2 \phi \,
	\frac{\e{\sqrt{\lambda}\cos\phi} +\e{\sqrt{\lambda}\cos\theta}}{\e{\sqrt{\lambda}\cos\phi}-\e{\sqrt{\lambda}\cos\theta}}~.
\end{equation}
Here, the $\phi$-integral must be interpreted as the principle value because of the pole for $\phi=\theta$, whereas there was no pole with the symmetric integrand. However, after rewriting \eqref{WL1:integ.C2.1} as 
\begin{equation}
\label{WL1:integ.C2.2}
	C_2 = -\frac{\sqrt{\lambda}}{4\pi^2} \int\limits_0^{\ths} \rmd \theta \int\limits_0^{\ths} \rmd \phi \sin \phi \,
	\partial_\phi \left[ \ln \left|\e{\sqrt{\lambda}(\cos\theta-\cos\phi)} - 1 \right| + 
		\ln \left|\e{\sqrt{\lambda}\cos\phi} - \e{\sqrt{\lambda}\cos\theta} \right| \right]~,
\end{equation}
the $\phi$-integral is easily done using integration by parts, which also takes care of the principal value. The result is 
\begin{equation}
\label{WL1:integ.C2}
	C_2 = \frac{\lambda}{8\pi^2} \left( -\sin^2 \ths + \frac12 \ths \sin 2 \ths \right)~.
\end{equation}
Summing the results \eqref{WL1:integ.A}, \eqref{WL1:integ.Ap}, \eqref{WL1:integ.B}, \eqref{WL1:integ.C1} and \eqref{WL1:integ.C2}, we obtain the first line of \eqref{WL1:FF1.integ},
\begin{equation}
\label{WL1:tFF1.holo}
	\widetilde{\FF[1]}(t) = \frac{\lambda}{8\pi^2} \left( \ths^2 - \ths \sin 2 \ths + \sin^2 \ths \right)~.
\end{equation}
This agrees with (3.53b) of \cite{Gordon:2017dvy}.\footnote{The difference in the sign stems from the different definitions of $\FF(t)$.}  

Finally, after adding \eqref{WL1:tFF0.holo} to \eqref{WL1:tFF1.holo} and using \eqref{WL1:kappa}, we end up with the final result for $\FF[1](t)$, 
\begin{equation}
\label{WL1:FF1.holo}
	\FF[1](t) = \frac{\lambda}{8} \left[ -\kappa (1-\kappa) + \frac{1}{\pi^2} \sin^4 \ths \right]~.
\end{equation}
We note that \eqref{WL1:FF1.holo} is symmetric under $\ths\to \pi -\ths$, which lets $\kappa\to 1-\kappa$. 
This reflects, of course, the palindromic property of $F_A(t;I(g,g))$. We also note that the first term in brackets in \eqref{WL1:FF1.holo} cancels against the exponential factor in \eqref{GF:WL.SUN.simple}, \ie for the $SU(N)$ Wilson loop. This reproduces (3.56) of \cite{Gordon:2017dvy}.

\section{Conclusions}\label{concs}
In this paper we tackled the problem of non-planar corrections to the expectation value of $\frac{1}{2}$-BPS Wilson loops in $\mathcal{N}=4$ SYM with gauge group $U(N)$ or $SU(N)$. More precisely, we extracted the leading and sub-leading behaviours in the $1/N$ expansion at fixed 't~Hooft coupling $\lambda$ of the Wilson loop generating function. Unlike previous works, which had addressed this issue using loop equation techniques and topological recursion, our starting point was the exact solution of the matrix model, which had been known for some time. Our results for the $1/N$ term of the Wilson loop generating function agree with previous calculations, but appear to be somewhat more explicit. We have provided both sum and integral representations of the $1/N$ terms and have evaluated them explicitly in the holographic large-$\lambda$ regime, which allows for easier comparison with the holographic dual picture. This term should match with the gravitational backreaction of the D-brane on the gravity side.
A particularly interesting observation is the connection between the Wilson loop generating function and the finite-dimensional quantum system known as the truncated harmonic oscillator. This system, which is familiar to the Quantum Optics community, provides a description of the problem that seems to be more amenable to an asymptotic $1/N$ expansion.
En route, we obtained interesting mathematical relations and sum rules involving the Hermite polynomials. It would be interesting to see these formulas, which we proved in appendices~\ref{HeNU} and \ref{SA}, in different applications. 

One can envisage two main lines of generalization of the present work. First, it would be interesting to extend the methods developed here to other representations of the gauge group. A particularly interesting case is the totally symmetric representation $\mathcal{S}_k$ (see also \cite{Chen-Lin:2016kkk}), whose generating function is slightly more complicated than the antisymmetric one but still quite simple, so the problem seems tractable. 
Second, one may investigate how the present approach extends to higher orders in $1/N$. At first sight, there are a number of technical obstacles that must be overcome, because order-$1/N^2$ terms have been neglected at many points of the calculation. So, the question whether our approach lends itself to a systematic $1/N$ expansion is highly non-trivial. This problem is closely related to the fact that the large, but finite-$N$ matrix model solution differs in its analyticity properties from the continuum limit. Moreover, corrections to the saddle point calculation of the Wilson loop expectation values become relevant at order $1/N^2$. We leave these interesting questions for the future.

\section*{Acknowledgements}
A.C.\ and A.F.\ were supported by Fondecyt \# 1160282. The research of W.M.\ was partly supported by the I.N.F.N., research initiative STEFI.

\begin{appendix}
\section{Some properties of the Hermite polynomials}
\label{Hermite}

In this appendix, we list a number of formulae regarding the (probabilists') Hermite polynomials, which are useful for the analysis in the main text. These relations can be found in standard references \cite{Gradshteyn, NIST}. Sometimes a translation from the physicists' version of the polynomials is necessary. They are related by
\begin{empheq}{equation}
\label{HeN:polys}
	\He_n(x) = 2^{-\frac{n}{2}} \Hermite_n(x/\sqrt{2})~.
\end{empheq}

The Hermite polynomials $\He_n(x)$ satisfy the differential equation
\begin{empheq}{equation}
\label{HeN:deq}
	\He_n'' - x \He_n' + n \He_n = 0 
\end{empheq}
as well as the recurrence relations
\begin{empheq}{align}
\label{HeN:rec1}
	\He_n' &= n \He_{n-1}~,\\
\label{HeN:rec}
	\He_{n+1} &=  x \He_n - n \He_{n-1}~. 
\end{empheq}
The generating function is 
\begin{empheq}{equation}
\label{HeN:gen.func}
	\e{xt-\frac12 t^2} = \sum\limits_{n=0}^\infty \frac{t^n}{n!} \He_n(x)~.
\end{empheq}
Another useful property is the linearization formula
\begin{empheq}{equation}
\label{HeN:lin.form}
	\He_m(x) \He_n(x) = \sum\limits_{k=0}^m \binom{m}{k}
	\binom{n}{k} k! \He_{m+n-2k}(x)~.
\end{empheq}

A main ingredient in our analysis is the location of the $N$ roots of $\He_N$ for large $N$. It can be obtained from the relation of $\He_N$ to the parabolic cylinder function
\begin{empheq}{equation}
\label{HeN:par.cyl}
	\He_N(x) = \e{x^2/4} \operatorname{U} \left(-N-\frac12, x\right)
\end{empheq}
and the asymptotic expansion of the parabolic cylinder function 
\begin{empheq}{equation}
\label{HeN:par.cyl.asympt}
	\operatorname{U} \left(-\frac12 \mu^2, \sqrt{2}\mu t\right) \sim \frac{2 g(\mu)}{(1-t^2)^{1/4}} 
	\left[ \cos\kappa \sum\limits_{s=0}^\infty (-1)^s 
		\frac{u_{2s}(t)}{(1-t^2)^{3s} \mu^{4s}} - \sin \kappa \sum\limits_{s=0}^\infty (-1)^s 
		\frac{u_{2s+1}(t)}{(1-t^2)^{3s+3/2} \mu^{4s+2}} \right]~,
\end{empheq}
where
\begin{empheq}{equation}
\label{HeN:kappa}
	\kappa = \mu^2 \eta -\frac14 \pi~, \qquad \eta = \frac12 \left( \arccos t - t \sqrt{1-t^2} \right)~,
\end{empheq}
and $u_s(t)$ are polynomials
\begin{empheq}{equation}
\label{HeN:un}
	u_0= 1, \qquad u_1 = \frac1{24} t(t^2-6)~, \quad \cdots
\end{empheq}
The function $g(\mu)$ is irrelevant for our purposes. Setting $\mu = \sqrt{2N+1}$ and $t= \cos\theta$, these relations imply that the $N$ roots of $\He_N$ are given approximately by 
\begin{empheq}{equation}
\label{HeN:roots1}
	\zeta_i = 2\sqrt{N+\frac12}\cos\theta_i~,
\end{empheq} 
with
\begin{empheq}{equation}
\label{HeN:roots2}
	\left( N+\frac12 \right) \left( \theta_i - \frac12 \sin 2 \theta_i \right) - \frac14 \pi = \left( i-\frac12\right) \pi
	\qquad ( i = 1, 2, \ldots, N)~. 
\end{empheq}

\section{Trace of normal ordered products}
\label{calc}

In this appendix, we calculate the trace of the normal ordered product $A^T{}^m A^n$.
First, consider $A^T{}^n A^n$. Using the definition of the number operator \eqref{MM:N.def} and the commutators \eqref{MM:N.A.comm}, we get
\begin{equation}
\label{calc:AAn}
	A^T{}^n A^n = A^T{}^{n-1} \N A^{n-1} = A^T{}^{n-1} A^{n-1} (\N -n+1) = (\N -n+1)_n~, 
\end{equation}
where we have iterated the first two steps to arive at the final expression. $(a)_n=\Gamma(a+n)/\Gamma(a)$ denotes the Pochhammer symbol. This allows us to write immediately
\begin{equation}
\label{calc:AmAn}
	A^T{}^m A^n = 
		\begin{cases}
			A^T{}^{m-n} (\N -n+1)_n \quad &\text{for $m\geq n$,}\\
			(\N -m+1)_m A^{n-m} &\text{for $m< n$.}
		\end{cases}		
\end{equation}
Because $\N$ is diagonal and any power of $A$ is off-diagonal, the trace of this quantity is non-zero only, if $m=n$,
\begin{equation}
\label{calc:AmAn.tr}
	\Tr\left( A^T{}^m A^n \right) = \delta_{mn} \Tr (\N -n+1)_n~.
\end{equation}
The trace in \eqref{calc:AmAn.tr} can be calculated starting with the expansion of the Pochhammer symbol in terms of Stirling numbers \cite{NIST}. This gives
\begin{align}
\notag
	\Tr (\N -n+1)_n &= \sum_{l=0}^n s(n,l) \Tr \N^l  
	= \sum_{l=0}^n s(n,l) \sum_{k=0}^{N-1} k^l \\
	&= \sum_{l=1}^n s(n,l) \sum_{j=0}^l j! \binom{N}{j+1} S(l,j)~,	
\label{calc:AAn.tr1}
\end{align}
where $S(l,j)$ denote the Stirling numbers of the second kind. 
After rearranging the sum and using the properties of the Stirling numbers, \eqref{calc:AAn.tr1} becomes 
\begin{equation}
\label{calc:AAn.tr2}
	\Tr (\N -n+1)_n = n! \binom{N}{n+1} = \frac{(N-n)_{n+1}}{n+1}~.
\end{equation}
Finally, expressing the Pochhammer symbol in terms of Stirling numbers of the first kind yields an expansion in $1/N$,
\begin{equation}
\label{calc:AAn.tr3}
	\frac1{N^{n+1}} \Tr (\N -n+1)_n = \frac1{n+1} \sum_{l=0}^n s(n+1,n+1-l) N^{-l}~.
\end{equation}

\section{Conversion of the sum over the roots of $\He_N$ to an integral}
\label{SI}

Consider a sum of the form
\begin{empheq}{equation}
	\frac{1}{N}\sum_{i=1}^Nf(\theta_i)\,,
\end{empheq}
where $\theta_i$ are defined in \eqref{HeN:roots1} in terms of the zeros of the Hermite polynomial $\textrm{He}_N(\zeta)$. In the large $N$ limit, it is justified to convert such a sum into an integral. Here we describe here how to do it correctly to order $1/N$. 

We start by setting $x=i-\frac12$ and use the Euler-Maclaurin formula in mid-point form \cite{Sarafyan:1979}. The mid-point form has the advantages that the integration domain lies manifestly symmetric within the inveral $(0,N)$ and that the boundary terms in the Euler-Maclaurin formula contain only derivatives of the integrand. The latter turn out to contribute at least of order $1/N^2$ and are, therefore, irrelevant for our purposes. Hence, we have
\begin{empheq}{align}
\notag 
	\frac1N \sum_{i=1}^{N} f(\theta_i) &= \frac1N \int\limits_0^N \rmd x f(\theta_{x+1/2}) +  \Order\left(\frac1{N^2}\right) \\
\label{SI:Euler.ML}
	& = 
	\frac2{\pi} \left(1+\frac1{2N} \right) \int\limits_{\theta_{1/2}}^{\pi-\theta_{1/2}} \rmd \theta \sin^2 \theta f(\theta)
	+ \Order\left(\frac1{N^2}\right)\,.
\end{empheq}
In the second equality, we have changed the integration variable using \eqref{HeN:roots1}. The angle $\theta_{1/2}$, which marks the tiny edges missing from the interval $(0,\pi)$, also follows from \eqref{HeN:roots1},
\begin{empheq}{equation}
\label{SI:theta.int.limit}
	\left( N+\frac12 \right) \left( \theta_{1/2} - \frac12 \sin 2 \theta_{1/2} \right) = \frac14 \pi \qquad 
	\Rightarrow \qquad \theta_{1/2}^3 \approx \frac{3\pi}{8N}~.
\end{empheq}
Therefore, \eqref{SI:Euler.ML} becomes
\begin{empheq}{equation}
\label{SI:Euler.ML2}
	\frac1N \sum_{i=1}^{N} f(\theta_i) =  
	\frac2{\pi} \left(1+\frac1{2N} \right) \int\limits_0^\pi \rmd \theta \sin^2 \theta f(\theta) 
	-\frac1{4N} \left[ f(0) + f(\pi) \right]+ \Order\left(\frac1{N^2}\right)~.
\end{empheq}
To obtain the second term on the right hand side we have assumed that the function $f$ is regular at $0$ and $\pi$. 

Let us check the formula \eqref{SI:Euler.ML2} by calculating the trace of the matrix $I(g,g)$ in the position basis \eqref{WL1:mat.el4} and compare the result with the exact expression, which can be calculated in the number basis as follows,\footnote{This calculation appeared already in section~\ref{WL0} between \eqref{WL0:F1} and \eqref{WL0:F1.N01} and makes use of the calculation in appendix~\ref{calc}.}
\begin{align}
\notag 
	\frac1N \Tr \left( \e{gA^\dagger} \e{gA} \right) &=
	\sum\limits_{k,l=0}^{N-1} \frac{g^{k+l}}{k!l!} \frac1N \Tr \left( A^\dagger{}^k A^l \right) \\
\notag
	&= \sum\limits_{k=0}^{N-1} \frac{g^{2k}}{(k!)^2} \frac1N \Tr (\N -k+1)_k \\
\notag
	&= \sum\limits_{k=0}^{N-1} \frac{g^{2k}N^k}{k!(k+1)!} \sum\limits_{l=0}^k s(k+1, k+1-l) N^{-l} \\
\label{WL1:trace.exact}
	&= \left( 1-\frac{\lambda}{8N} \right) \frac2{\sqrt{\lambda}} \BesselI[1](\sqrt{\lambda}) 
	+\Order \left(\frac1{N^2}\right)~.	
\end{align}

In the position basis, we have from \eqref{WL1:mat.el4} and \eqref{SI:Euler.ML2}
\begin{align}
\notag
	\frac1N \Tr I(g,g) &=
	\frac2{\pi} \left(1+\frac1{2N} \right) \int\limits_0^\pi \rmd \theta \sin^2 \theta 
	\e{2g\sqrt{N+\frac12} \cos\theta-\frac12 g^2}  \\
\notag
	&\quad +\frac1N \sum\limits_{r,s=1}^\infty \BesselI[r+s](\sqrt{\lambda}) 
	\frac{2}{\pi} \int\limits_0^\pi \rmd \theta \sin(r\theta) \sin(s\theta)  - \frac1{2N} \cosh \sqrt{\lambda} \\
\notag
	&= \left(1+\frac1{2N} \right) \e{-\frac{\lambda}{8N}} \frac2{\sqrt{\lambda}{\sqrt{1+\frac1{2N}}}} \BesselI[1] 
	\left(\sqrt{\lambda} \sqrt{1+\frac1{2N}}\right) \\
\notag
	&+ \frac1N \sum_{r=1}^\infty \BesselI[2r](\sqrt{\lambda}) - \frac1{2N} \cosh \sqrt{\lambda}\\
\notag 
	&= \frac2{\sqrt{\lambda}} \BesselI[1](\sqrt{\lambda}) \e{-\frac{\lambda}{8N}} 
	+ \frac1N \left[ \frac12 \BesselI[0](\sqrt{\lambda}) +  \sum_{r=1}^\infty \BesselI[2r](\sqrt{\lambda}) 
	-\frac12 \cosh\sqrt{\lambda} \right]\\
\label{WL1:trace.check}
	&= \frac2{\sqrt{\lambda}} \BesselI[1](\sqrt{\lambda}) \e{-\frac{\lambda}{8N}}~.
\end{align}
We especially point out the presence of the last term on the second line, which comes from the edge terms of \eqref{SI:Euler.ML2} and is crucial for cancelling other $1/N$ contributions. The bracket on the penultimate line vanishes by means of a well-known summation formula of the modified Bessel functions \cite{NIST}. In these expressions, we have dropped all contributions of order $1/N^2$. Obviously \eqref{WL1:trace.check} agrees with \eqref{WL1:trace.exact} to order $1/N$.

\section{Proof of \eqref{WL1:Hermite.Chebychev}}
\label{HeNU}

In this appendix, we provide a proof of the asymptotic formula
\begin{equation}
\label{HeNU:Hermite.Chebychev}
	\frac{\He_{N-s}(\zeta)}{\He_{N+1}(\zeta)} \sim - \frac{(N-s)!}{N!} N^{\frac{s-1}2} \ChebU_{s-1}(\cos \theta)~,
\end{equation}
which is \eqref{WL1:Hermite.Chebychev} in the main text. This formula holds for $s\ll N$, and $\zeta \sim 2\sqrt{N}\cos \theta$ is a root of $\He_N$. 

The proof is done by induction using the recursion formula for the Hermite polynomials \eqref{HeN:rec}. First, consider $s=1$ and $s=2$. Because $\zeta$ is a root of $\He_N$, we have
\begin{align}
\label{HeNU:1}
	\He_{N-1}(\zeta) &= -\frac1N \He_{N+1}(\zeta)~, \\
\label{HeNU:2}
	\He_{N-2}(\zeta) &= \frac1{N-1} \zeta \He_{N-1}(\zeta) = - \frac1{N(N-1)} N^{\frac12} 2 \cos \theta \He_{N+1}(\zeta)~,
\end{align}
so \eqref{HeNU:Hermite.Chebychev} obviously holds for $s=1$ and $s=2$. Now, assume that \eqref{HeNU:Hermite.Chebychev} holds for $\He_{N-s+1}$. Then, from the recursion formula \eqref{HeN:rec} we get
\begin{align}
\notag
	\He_{N-s}(\zeta) &= \frac1{N-s+1} \left[ \zeta \He_{N-s+1}(\zeta) - \He_{N-s+2}(\zeta) \right]~, \\
\notag
	\frac{\He_{N-s}(\zeta)}{\He_{N-1}(\zeta)} &= 
		- \frac1{N-s+1} \left[ 
		 2\sqrt{N} \cos\theta \frac{(N-s+1)!}{N!} N^{\frac{s-2}2}\ChebU_{s-2}(\cos\theta)  \right. \\
\notag 
	&\quad \left. - \frac{(N-s+2)!}{N!}N^{\frac{s-3}2}\ChebU_{s-3}(\cos\theta) \right] \\
\notag
	&= - \frac{(N-s)!}{N!} N^{\frac{s-1}2} \left[ 2\cos\theta \ChebU_{s-2}(\cos\theta) 
		- \frac{N-s+2}{N} \ChebU_{s-3}(\cos\theta) \right] \\
\notag
	&\sim - \frac{(N-s)!}{N!} N^{\frac{s-1}2} \ChebU_{s-1}(\cos\theta)~,
\end{align}
where we have dropped the term s of order $1/N$ in front of $\ChebU_{s-3}(\cos\theta)$ and used the recursion relation for the Chebychev polynomials \cite{Gradshteyn} in the last step.

\section{Proof of \eqref{WL1:Sa}}
\label{SA}

In this appendix, we shall prove the remarkable summation formula 
\begin{equation}
\label{SA:Sa}
	\sum\limits_{r=1}^v \sum\limits_{t_1=1}^\infty  \cdots \sum\limits_{t_{a-1}=1}^\infty 
		 	\BesselI[v-r+t_1](z) \BesselI[t_2-t_1](z) \cdots \BesselI[r-t_{a-1}](z) = \frac{v}{a} \BesselI[v](az)\,,
		 	\qquad a = 2,3,\ldots\,.
\end{equation}
For $a=2$, we have established the result in \eqref{WL1:S2}. Our starting point is the generating function of the modified Bessel functions \eqref{WL1:Bessel.gen},
\begin{equation}
\label{SA:Bessel.gen}
	\e{z\cos\theta} =  \BesselI[0](z) + 2 \sum\limits_{k=1}^\infty \BesselI[k](z) \cos(k\theta)~.
\end{equation}
Differentiating it with respect to $\theta$ shows that 
\begin{equation}
\label{SA:Bessel.gen2}
	\frac{z}2 \sin \theta \e{az\cos\theta} = \sum\limits_{k=1}^\infty \frac{k\BesselI[k](az)}{a} \sin(k\theta)~.
\end{equation}
Therefore, we can prove \eqref{SA:Sa} by showing that 
\begin{equation}
\label{SA:Sa.gen}
	\sum\limits_{v=1}^\infty S_a(v;z) \sin(v\theta)= \frac{z}2 \sin \theta \e{az\cos\theta}\,,
\end{equation}
where $S_a(v;z)$ is, as defined in the main text, the left hand side of \eqref{SA:Sa}. Since $S_2(v;z)$ is known from \eqref{WL1:S2}, \eqref{SA:Sa.gen} trivially holds for $a=2$ by virtue of \eqref{SA:Bessel.gen2}.

The left hand side of \eqref{SA:Sa.gen} is explicitly 
\begin{equation}
\label{SA:Sa.calc}
	\sum\limits_{v=1}^\infty S_a(v;z) \sin(v\theta) 
	= \sum\limits_{v=1}^\infty \sin(v\theta) \sum\limits_{r=1}^v \sum\limits_{t_1=1}^\infty 
		\cdots \sum\limits_{t_{a-1}=1}^\infty 
			\BesselI[v-r+t_1](z) \BesselI[t_2-t_1](z) \cdots \BesselI[r-t_{a-1}](z)\,.
\end{equation}
We rearrange the sums over $v$ and $r$ by introducing $s=v-r$ and $t_a=r$, such that
\begin{equation}
\label{SA:Sa.calc2}
	\sum\limits_{v=1}^\infty S_a(v;z) \sin(v\theta) 
	= \sum\limits_{s=0}^\infty \sum\limits_{t_a=1}^\infty 
		\cdots \sum\limits_{t_1=1}^\infty \sin[(s+t_a)\theta] \BesselI[t_a-t_{a-1}](z) \cdots \BesselI[t_2-t_1](z) 
		\BesselI[s+t_1](z)\,.
\end{equation}
Now, we can extend the summation over $t_a$ to all integers without changing the value of the sum. The reason for this is that
\begin{equation}
	\sum\limits_{s=0}^\infty \sum\limits_{r=0}^\infty \sin[(s-r)\theta] M_{rs} =0~,
\end{equation}
if $M_{rs}$ is symmetric, by the antisymmetry of the sine. It is easy to see that this is the case for the matrix involving the rest of the $t$-summations and the modified Bessel functions in \eqref{SA:Sa.calc2}. Therefore,
\begin{align}
\notag
	\sum\limits_{v=1}^\infty S_a(v;z) \sin(v\theta) 
	&= \sum\limits_{s=0}^\infty \sum\limits_{t_a=-\infty}^\infty \sum\limits_{t_{a-1}=1}^\infty 
		\cdots \sum\limits_{t_1=1}^\infty \sin[(s+t_a)\theta] \BesselI[t_a-t_{a-1}](z) \cdots \BesselI[t_2-t_1](z) 
		\BesselI[s+t_1](z)\\
\notag
	&= \sum\limits_{s=0}^\infty \sum\limits_{t_{a-1}=1}^\infty 
		\cdots \sum\limits_{t_1=1}^\infty \sum\limits_{t_a=-\infty}^\infty 
		\left\{\sin[(s+t_{a-1})\theta] \cos(t_a \theta) + \cos[(s+t_{a-1})\theta] \sin(t_a \theta) \right\} \\
\notag &\quad \times \BesselI[t_a](z) 
		\BesselI[t_{a-1}-t_{a-2}](z) \cdots \BesselI[t_2-t_1](z) 
		\BesselI[s+t_1](z)\\
	&= \e{z\cos\theta} \sum\limits_{v=1}^\infty S_{a-1}(v;z) \sin(v\theta)~.
\end{align}
In the step to the last line, we have carried out the sum over $t_a$ using \eqref{SA:Bessel.gen} and used the fact that the sum of $\sin(t_a\theta) \BesselI[t_a](z)$ vanishes. Continuing recursively, we obtain
\begin{equation}
\label{SA:Sa.proof}
	\sum\limits_{v=1}^\infty S_a(v;z) \sin(v\theta) 
	= \e{(a-2)z\cos\theta} \sum\limits_{v=1}^\infty S_{2}(v;z) \sin(v\theta)
	= \frac{z}2 \sin \theta \e{az\cos\theta}~,
\end{equation}
knowing that \eqref{SA:Sa.gen} is true for $a=2$. Thus, we have proven \eqref{SA:Sa.gen} for all $a\geq2$, which implies \eqref{SA:Sa}. 

\end{appendix}


\providecommand{\href}[2]{#2}\begingroup\raggedright\endgroup

\end{document}